\documentclass[11pt,a4paper,english,twoside]{article}

\usepackage{a4wide}
\usepackage{amssymb, amsmath}
\usepackage{graphicx}
\usepackage[all]{xy}
\usepackage{hyperref}
\hypersetup{linktocpage}
\usepackage{enumerate}
\usepackage{xcolor}
\usepackage{dsfont}
\usepackage{empheq}
\usepackage{cite}
\usepackage{float}

\newcommand{\beq}{\begin{equation}}
\newcommand{\eeq}{\end{equation}}
\def\bea#1\eea{\begin{align}#1\end{align}}
\def\beal#1\eeal{\begin{subequations}\begin{align}#1\end{align}\end{subequations}}
\newcommand{\nn}{\nonumber}
\newcommand{\w}{\wedge}

\def\del {\partial}
\def\d {{\rm d}}
\def\mmm {\mathcal{M}}

\begin{document}
\numberwithin{equation}{section}

\begin{titlepage}

\begin{flushright}
CERN-TH-2017-222
\end{flushright}

\begin{center}

\phantom{DRAFT}

\vspace{1.5cm}

{\LARGE \bf{On classical de Sitter and Minkowski solutions \vspace{0.4cm}\\ with intersecting branes}}\\

\vspace{2 cm} {\Large David Andriot}\\
 \vspace{0.9 cm} {\small\slshape CERN, Theoretical Physics Department\\
1211 Geneva 23, Switzerland}\\
 \vspace{0.5cm} {\upshape\ttfamily david.andriot@cern.ch}\\

\vspace{3cm}

{\bf Abstract}
\vspace{0.1cm}

\end{center}

\begin{quotation}
Motivated by the connection of string theory to cosmology or particle physics, we study solutions of type II supergravities having a four-dimensional de Sitter or Minkowski space-time, with intersecting $D_p$-branes and orientifold $O_p$-planes. Only few such solutions are known, and we aim at a better characterisation. Modulo a few restrictions, we prove that there exists no classical de Sitter solution for any combination of $D_3$/$O_3$ and $D_7$/$O_7$, while we derive interesting constraints for intersecting $D_5$/$O_5$ or $D_6$/$O_6$, or combinations of $D_4$/$O_4$ and $D_8$/$O_8$. Concerning classical Minkowski solutions, we understand some typical features, and propose a solution ansatz. Overall, a central information appears to be the way intersecting $D_p$/$O_p$ overlap each other, a point we focus on.
\end{quotation}

\end{titlepage}

\newpage

\noindent\rule[1ex]{\textwidth}{1pt}
\vspace{-0.8cm}

\tableofcontents

\vspace{0.5cm}
\noindent\rule[1ex]{\textwidth}{1pt}
\vspace{0.4cm}

\section{Introduction}

String theory is a fascinating quantum gravity theory that contains all necessary ingredients to be a fundamental theory of high energy physics. But connecting it to real, observable, physics remains so far out of reach. A major difficulty in such a relation lies in the richness of string theory: it has several features that are unobserved, but contribute crucially to the path to quantum gravity; mechanisms should then be found to explain why they are not detected. Two important examples are the extra dimensions, that naturally address the hierarchy problem, or supersymmetry, that plays a crucial role in U.V.~finiteness. Accommodating these two unobserved features will also be a challenge in the present work. Here, we are mainly interested in the connection to cosmology. To address this question, we study the existence of de Sitter solutions of type II supergravities.

\newpage

\begin{figure}[H]
\begin{center}
\includegraphics[width=6.0cm]{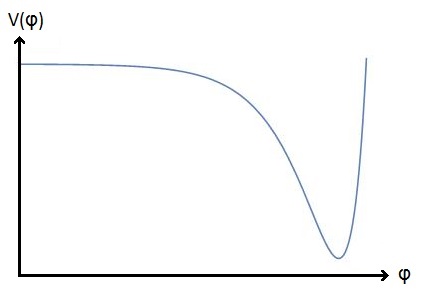}\caption{Typical potential of a single field inflation model in agreement with observational constraints. The end-point is a stable de Sitter solution, while the inflation phase is almost an (unstable) de Sitter solution.}\label{fig:pot}
\end{center}
\end{figure}

\begin{itemize}
\item {\bf Connecting cosmology to string theory}
\end{itemize}
The recent cosmological observations \cite{Ade:2015tva, Ade:2015xua, Ade:2015lrj} at high precision have brought important constraints on the description of the early universe. While several cosmological models have been ruled-out, many others are however still allowed. In addition, their embedding into more complete theories, such as four-dimensional supergravities, is often realised. It would thus be interesting for cosmology to have theoretical criteria allowing to distinguish between these various models. An important criterion would be the realisation of the model in a quantum gravity theory (see \cite{Barrau:2017tcd} for a recent review), such as string theory. This would provide in principle a U.V.~description. From the string theory side, the connection to cosmology is certainly required, but it is also a particulary interesting area to establish a relation to observable physics. Indeed, contrary to the usual landscape idea, a well-controlled connection between string theory and a cosmological model is difficult to establish, and could in the end be very special, non-generic, if not unique.
\begin{itemize}
\item {\bf Why (metastable) de Sitter solutions?}
\end{itemize}
To tackle the connection to cosmology, we focus on the question of de Sitter solutions: those admit a four-dimensional de Sitter space-time, i.e.~with positive cosmological constant $\Lambda$ or four-dimensional Ricci scalar ${\cal R}_4=4\Lambda >0$. We first recall that a spatially flat FLRW metric with an exponential scale factor $a(t) = e^{Ht}$ corresponds to a de Sitter space-time, with $\Lambda=3 H^2$. If one describes the early universe with an inflation model, there are three points or phases in the universe evolution that are close to having a four-dimensional de Sitter space-time: first, the present universe, that is attracted towards a pure de Sitter solution (as long as the observed $\Lambda$ is constant); second, the end-point of inflation, which is a minimum of the inflaton potential $V(\varphi)$ with typically a positive value $V=2\Lambda$; third, the inflation phase itself, for slow-roll models, is almost a de Sitter solution, since $V(\varphi)$ is then positive and almost flat: see Figure \ref{fig:pot}. Therefore, even though having a four-dimensional de Sitter space-time will not describe the entire evolution of the universe, it could be used to match one of these three points, as a fixed point or static limit, and serve this way as a stepping stone to build a more complete model.
\vspace{-0.2cm}
The stability of a de Sitter solution is another important aspect. The end-point of inflation is a solution, meaning an extremum $\del_{\varphi} V =0$, but also a minimum or vacuum, $\del^2_{\varphi} V >0$, i.e.~the solution is metastable if not stable. This is commonly required for the reheating process to happen, through inflaton oscillations. The inflation phase in a slow-roll model is almost a de Sitter solution, slightly unstable. Finally, the de Sitter fixed point towards which our present universe is attracted should also be (meta)stable. Therefore, one usually looks for metastable de Sitter solutions; if they are rather found tachyonic, one can still compare the potential $\eta$-parameter to that of a slow-roll inflation. Here, we focus on the existence of de Sitter solutions, and postpone to future work the study of stability.
\begin{itemize}
\item {\bf (Classical) de Sitter and Minkowski solutions from string theory}
\end{itemize}
To connect string theory to a four-dimensional model and accommodate the extra dimensions, one usually works in the context of a compactification. The starting point is a theory in a ten-dimensional space-time. One then requires to have maximally symmetric four-dimensional space-time (e.g.~de Sitter or Minkowski), so the ten-dimensional one is split as a (warped) product of the four-dimensional space-time and a six-dimensional compact (internal) manifold $\mmm$. One looks for a solution of this form to the ten-dimensional equations of motion, and possible other constraints. Given this solution, one can perform a dimensional reduction, resulting in a four-dimensional theory with a scalar potential (e.g.~that of inflation): the extremum of the latter should correspond to the ten-dimensional solution. In this context, a de Sitter solution can be obtained in different manners (see \cite{Bena:2017uuz} for a recent review). First, one can consider different theories in ten dimensions, namely the various string theories, their low energy supergravities, or further approximations thereof (heterotic, F-theory, etc.), and look at that level for a solution. If the theory is a supergravity without higher order stringy corrections, the solution is said to be classical. One may also work with a four-dimensional theory and find a solution by studying the potential. In that case, one could obtain a solution with $V>0$ either directly as a classical de Sitter solution, or with a Minkowski (or even anti-de Sitter) solution at tree level that gets higher order or even non-perturbative, positive, corrections \cite{Kachru:2003aw, Balasubramanian:2005zx, Westphal:2006tn}. In this last case, the question is then whether the corrections to the potential can be embedded, with the classical solution, into a consistent ten-dimensional picture, or whether the four-dimensional theory lies in the swampland. We refer to \cite{Bena:2009xk, Cohen-Maldonado:2015ssa, Bena:2016fqp, deAlwis:2016cty, CaboBizet:2016qsa, Moritz:2017xto, Sethi:2017phn} for various discussions on such constructions, and to \cite{Cicoli:2017shd, Gallego:2017dvd} for recent examples.

Here, we work in ten-dimensional type IIA or IIB supergravity with $D_p$-branes and orientifold $O_p$-planes, as a low energy effective theory of string theory. In this context, we focus on classical solutions with four-dimensional de Sitter or Minkowski space-time; those are then classical string backgrounds. We do not include higher order corrections in $\alpha'$ or the string coupling $g_s$, non-geometric or non-perturbative contributions, and do not allow for $N\!S_5$-branes or Kaluza-Klein monopoles. This framework, somehow restrictive, provides a good control on the relation between the ten-dimensional and four-dimensional pictures, which is important for a proper embedding of cosmological models. The (quantum) corrections to classical de Sitter solutions, and more generally the cosmological constant problem, will remain however to be studied if such solutions are found; those should depend on the precise solution. The motivation for classical de Sitter solutions is also to determine whether, as a matter of principle, such solutions can be found, before moving to more involved constructions.\footnote{Further motivations come from holography with the dS/CFT correspondence \cite{Hull:1998vg, Strominger:2001pn, Balasubramanian:2001nb}: see e.g.~\cite{Hertog:2017ymy, Neiman:2017zdr} for recent works on this topic. This is also related to the idea \cite{Verlinde:2016toy} of de Sitter space-time being an excited state with temperature, horizon and (entanglement) entropy, emerging from microscopic degrees of freedom.} The existence and stability of classical de Sitter solutions is an open question in type II supergravities, especially when one allows for all fluxes and a non-zero curvature of $\mmm$, i.e. so-called geometric fluxes.

The classical Minkowski solutions on the contrary can serve as a first background, to be further corrected towards a de Sitter solution. Minkowski solutions are also of major importance to realise particle physics models. Of particular interest here are the intersecting branes models: by considering intersecting stacks of $D_p$ and $O_p$, mostly with $p=6$ in type IIA supergravity on Minkowski times (an orbifold of) a torus, one can build a model that reproduces the particle physics standard model to some extent. We refer to \cite{Angelantonj:2002ct, Blumenhagen:2005mu, Blumenhagen:2006ci, Ibanez:2012zz, Honecker:2016gyz} for reviews. Providing examples of Minkowski solutions with intersecting branes could then be interesting for particle physics model building. In addition, we will look here for more involved solutions than those on a torus orbifold (see e.g.~\cite{Marchesano:2006ns, Berasaluce-Gonzalez:2016kqb, Berasaluce-Gonzalez:2017bib} for attempts of model building in this direction), allowing for fluxes and curved manifolds. Such a setup would help stabilizing closed string moduli, on top of the effects described in \cite{Blaszczyk:2015oia}, so it should be interesting for such constructions.
\begin{itemize}
\item {\bf Metastable classical de Sitter solutions: the status}
\end{itemize}
There is up-to-date no known metastable classical de Sitter string background. In view of the embedding of cosmological models into string theory, as explained above, this situation challenges the connection to cosmology. In heterotic string, de Sitter solutions have been ruled-out at all orders in $\alpha'$ and tree level in $g_s$ \cite{Green:2011cn, Gautason:2012tb, Kutasov:2015eba, Quigley:2015jia}. At higher order in the string coupling, the situation changes though, as indicated by the examples of \cite{Florakis:2016ani}, where the complete stability remains to be studied. Type II supergravities may then be the only framework where metastable classical de Sitter string backgrounds can be found. It remains a very difficult task, and many requirements or no-go theorems have been derived, starting with \cite{Gibbons:1984kp, deWit:1986xg, Maldacena:2000mw, Townsend:2003qv}, that are circumvented by including $O_p$. Many more works have refined this requirement \cite{Hertzberg:2007wc, Silverstein:2007ac, Covi:2008ea, Haque:2008jz, Caviezel:2008tf, Flauger:2008ad, Danielsson:2009ff, deCarlos:2009fq, Caviezel:2009tu, Wrase:2010ew, Shiu:2011zt, Burgess:2011rv, VanRiet:2011yc, Danielsson:2012et, Gautason:2013zw, Dasgupta:2014pma, Kallosh:2014oja, Junghans:2016uvg, Andriot:2016xvq, Junghans:2016abx}, often analysing a four-dimensional scalar potential, and studying the stability or the slow-roll inflation parameters. As an outcome, very few classical de Sitter solutions have been found \cite{Caviezel:2008tf, Flauger:2008ad, Danielsson:2009ff, Caviezel:2009tu, Danielsson:2010bc, Danielsson:2011au}, and none of them is metastable. In addition, no systematic origin of the observed tachyons has been discovered.

\begin{figure}[H]
\begin{center}
\includegraphics[width=7.0cm]{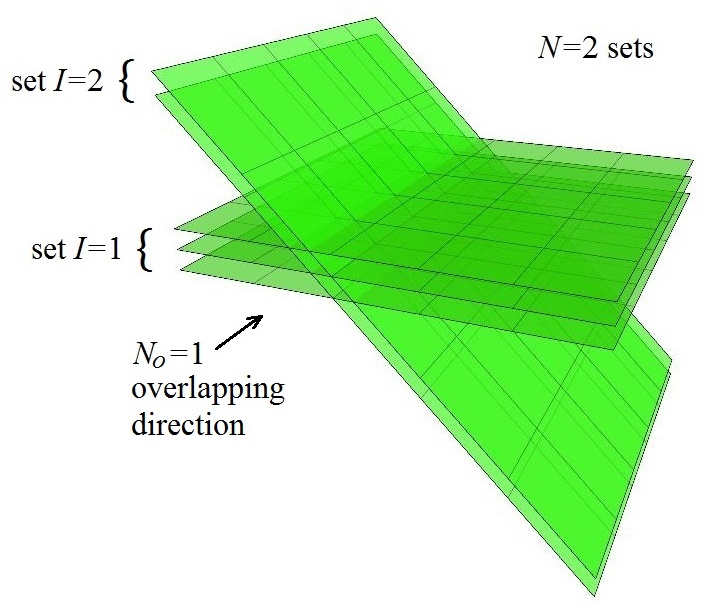}\caption{Each set $I=1,2$ is made of parallel $D_p/O_p$, and the different sets intersect each other. The $N=2$ sets have $N_o=1$ common (internal) direction, where their sources overlap.}\label{fig:branes}
\end{center}
\end{figure}

The constraints derived on classical de Sitter solutions are very dependent on the configuration of $D_p$ and $O_p$. These extended objects with $p+1$-dimensional world-volume may either be parallel or intersect each other: this distinction will play an important role in our analysis. Almost all classical de Sitter solutions found so far, all summarized in \cite{Danielsson:2011au}, were obtained on (orbifolds of) group manifolds and admit intersecting $O_6$ (see footnote \ref{foot:dSorbif} for details); the only exception being the solution of \cite{Caviezel:2009tu} that has both $O_5$ and $O_7$. These explicit examples will provide important checks of our results.
\begin{itemize}
\item {\bf The present work}
\end{itemize}
As motivated, we look for classical de Sitter or Minkowski solutions of type II supergravities with $D_p$ and $O_p$. For de Sitter solutions (and Minkowski with fluxes), the presence of $O_p$ is mandatory in this framework \cite{Maldacena:2000mw} while having $D_p$ is less important; for particle physics models however, having $D_p$ is crucial. We then refer to these sources collectively as $D_p/O_p$ without specifying the proportions of each type, but keeping in mind those constraints. We introduce a formalism to describe intersecting sources: for a fixed $p$, we consider $N$ sets of $D_p/O_p$ labeled by $I=1 \dots N$, where in each set the sources are parallel, but sources of sets $I\neq J$ are not; we then say that the latter intersect. Different sets may still have $N_o$ internal common directions, where the sources overlap. We summarize these notations with an example in Figure \ref{fig:branes}, and define them more precisely in Section \ref{sec:formal} and \ref{sec:homoover}. Finally, to preserve four-dimensional Lorentz invariance, we require the sources to fill the three extended space dimensions: we then restrict to $p\geq 3$ and further $p \leq 8$. Only the sources with $p \geq 4$ can then intersect, and do so in the internal manifold.
\vspace{-0.2cm}
In \cite{Andriot:2016xvq}, we obtained new and tight constraints on the existence of classical de Sitter solutions, in the case where the $D_p/O_p$ sources have a single size $p$ and are parallel, i.e.~$N=1$. Classical de Sitter solutions with $p=3$ were excluded in \cite{Blaback:2010sj}, building on \cite{Giddings:2001yu}; we showed in \cite{Andriot:2016xvq} that it was also case for parallel $p=7$ or $8$, outside of the F-theory regime. For parallel sources with $p=4,5,6$, tight constraints were obtained on a specific combination of internal curvature terms and fluxes. These results were derived up to minor assumptions on the sources and the internal manifold. Formal results of \cite{Andriot:2016xvq} were then used in \cite{Andriot:2016ufg} to find a class of classical Minkowski solutions with parallel $D_p/O_p$, extending \cite{Giddings:2001yu}. The present paper generalizes those studies to the case $N\geq 1$, i.e.~with intersecting $D_p/O_p$ sources, but also to sources of multiple sizes $p$. Having intersecting sources seems important for de Sitter solutions, given it is the case in the only known examples with single size $p$, while intersecting sources in Minkowski solutions could help building particle physics models, as explained previously.

The approach is analogous to that of \cite{Andriot:2016xvq, Andriot:2016ufg}. Having intersecting sources instead of parallel ones however adds several complications, and to start with, the backreaction, which is not considered in this paper. The method consists in deriving interesting expressions of ${\cal R}_4$ in terms of internal fields, by combining some equations of motion and Bianchi identities of the fluxes. The novelty, with respect to previous results, is the use of the trace of the Einstein equation along internal directions parallel to sources. For de Sitter solutions, the requirement ${\cal R}_4 >0$ then sets important constraints on the internal quantities, which can be turned into no-go theorems. For Minkowski solutions, imposing ${\cal R}_4 =0$ leads to some solution ansatz for the internal fields. In both cases, we make no use of supersymmetry, even though the knowledge of supersymmetric (Minkowski) solutions helps organising the fields and building interesting ${\cal R}_4$ expressions. Our results are summarized in Section \ref{sec:ccl}.

In more detail, the framework, conventions and useful equations are introduced in Section \ref{sec:formal}, supplemented with Appendix \ref{ap:Tmn}. We combine these equations in Section \ref{sec:deriv} to derive interesting ${\cal R}_4$ expressions, and obtain a no-go theorem for $p=7,8$ in Section \ref{sec:p=7,8}. Further constraints on de Sitter solutions are deduced in Section \ref{sec:nogo456} for $p=4,5,6$, with the interesting particular cases of homogeneous overlap discussed in Section \ref{sec:homoover} and specification to $\mmm$ being a group manifold in Section \ref{sec:groupmanif}. We turn to Minkowski solutions in Section \ref{sec:susymink}: the need for another ${\cal R}_4$ expression is motivated in Section \ref{sec:foreword}, it is derived in Section \ref{sec:derivcomment} and Appendix \ref{ap:HF}, and analysed in Section \ref{sec:Minksolfinal}. Finally, the case of sources of multiple sizes is studied in Section \ref{sec:multiplesizes}, leading in particular to a no-go theorem on de Sitter solutions for $p=3\, \&\, 7$. We summarize our results and give further comments in Section \ref{sec:ccl}.

\section{Formalities}\label{sec:formal}

We introduce in this section the framework, notations and equations we will need in the rest of the paper. We work in ten-dimensional (10d) type II supergravities, with $D_p$-branes and orientifold $O_p$-planes collectively referred to as sources. We consider no further ingredient. We follow the conventions detailed in \cite{Andriot:2016xvq}. The 10d space-time is split as a product of a 4d maximally symmetric space-time and a 6d internal compact manifold $\mmm$. The resulting metric is
\beq
\d s^2= g_{\mu\nu} (x) \d x^\mu \d x^\nu + g_{mn} (y) \d y^m \d y^n \ ,\label{10dmetric}
\eeq
where one would usually have in addition a warp factor, but we do not consider any here. The latter normally accounts for the backreaction of the sources, which we ignore in this paper. This can be understood as a smearing approximation, even though technically, the only thing we will do is to not consider any warp factor nor a varying dilaton. There are several reasons to be unsatisfied with this restriction, but it is a common one when studying intersecting sources as we will do. The solutions discussed can be viewed as a first step towards more complete ones or their stringy descriptions. The reason for this restriction is that no fully localized solution with intersecting branes is known in supergravity \cite{Smith:2002wn} (see also references in \cite{Macpherson:2016xwk}), with few exceptions \cite{Assel:2011xz, Assel:2012cj, Rota:2015aoa}. As a consequence, the dilaton $\phi$ will be considered constant from next section on. Therefore, our results do not cover F-theory type solutions. The remaining supergravity fields appear through the NSNS and RR fluxes: those will be captured by the purely internal forms $H$ and $F_{q=0\dots 6}$, as defined in \cite{Andriot:2016xvq}.

We turn to the sources. To preserve 4d Lorentz invariance, those fill the space of the 4d space-time, so we take $p\geq 3$, and further $p\leq 8$. As in \cite{Andriot:2016xvq}, we consider for simplicity that on each source $-\imath^*[b] + \mathcal{F} = 0$, where $\imath^*[\cdot]$ denotes the pull-back to the world-volume. Finally, we take for them $\mu_p=T_p$ as for BPS sources. We now turn to the embedding of the sources into the internal geometry. To describe it in practical terms, we make for each source the following geometric assumption on $\mmm$; as described in \cite{Andriot:2016xvq}, this is not very restrictive, as it includes at least fiber bundles. Working in the 6d flat (orthonormal) basis with metric $\delta_{ab}=e^m{}_a e^n{}_b g_{mn}$, we assume for each source the global separability of the 6d flat directions and one-forms $e^a$ into two sets, denoted $\{ e^{a_{||}} \}$ and $\{ e^{a_{\bot}} \}$. In mathematical terms, this amounts to a reduction of the structure group of the cotangent bundle from $O(6)$ to $O(p-3) \times O(9-p)$, or a subgroup thereof. Each one-form $e^{a_{||}}$ or $e^{a_{\bot}}$ does not need to be globally defined, only the separation is, and thus the two sets do not mix. Note that no assumption is made on the coordinate dependence. A more complete presentation is given in \cite{Andriot:2016xvq}. Wedging the one-forms of each set, one defines naturally internal parallel and transverse volume forms, and gets the relations
\bea
& {\rm vol}_{4} \w {\rm vol}_{||} \w {\rm vol}_{\bot} = {\rm vol}_{10} = \d^{10} x \sqrt{|g_{10}|} \ ,\label{volrel}\\
& {\rm vol}_{||} \w {\rm vol}_{\bot} = {\rm vol}_{6} = \d^{6} y \sqrt{|g_{6}|} \ ,\ *_6 {\rm vol}_{\bot} = (-1)^{9-p} {\rm vol}_{||} \ , \ *_6 {\rm vol}_{||} = {\rm vol}_{\bot} \ .\nn
\eea
To make contact with the source, we require its world-volume form to be given by
\beq
\d^{p+1} \xi \sqrt{|\imath^*[g_{10}]|}= \imath^*[{\rm vol}_{4} \w {\rm vol}_{||}] \ .\label{worldvolform}
\eeq
Finally, another requirement on the geometry will be needed, in some cases, at the end of the derivation when integrating, namely that one has for each source
\beq
f^{a_{\bot}}{}_{a_{\bot} b_{\bot}} = 0 \ ,\label{trace}
\eeq
with the definition $\d e^a = - \tfrac{1}{2} f^a{}_{bc} e^b \w e^c$. The condition \eqref{trace} is not always an assumption: it automatically holds if the transverse directions correspond to a smooth submanifold without boundary. Also, it is always satisfied with an orientifold on a group manifold where these $f^a{}_{bc}$ become structure constants: the compatibility of the orientifold projection with the algebra then sets $f^{a_{\bot}}{}_{b_{\bot} c_{\bot}}$ to zero.

In this work, we consider several, intersecting, sources. Each of them admits a split into its own $\{ e^{a_{||}} \}$ and $\{ e^{a_{\bot}} \}$. We do not require all these splits and one-forms to be defined in the same basis; rather, one may e.g.~have to rotate from one source to the other. In all sections but Section \ref{sec:multiplesizes}, we restrict ourselves to sources of a single fixed size $p$. This allows us to define parallel sources: those have the same directions $\{ e^{a_{||}} \}$ and $\{ e^{a_{\bot}} \}$, but could still be located at different points in their transverse space. Then, we consider $N$ different sets of parallel sources, labeled by $I=1 \dots N$: the sources in two different sets are not parallel, and are thus called intersecting. They may overlap along some directions, or not overlap at all. We summarize our notations with an example in Figure \ref{fig:branes}, where the quantity $N_o$ will be defined in Section \ref{sec:homoover}. Let us give another example: we consider $\mmm$ to be a flat torus with unit radii, various $O_5/D_5$ sources, and $N=3$ sets. The set $I=1$ admits $O_5/D_5$ along the internal $y^1$ and $y^2$, meaning that the $\{ e^{a_{||}} \}$ for this set is $\{\d y^1, \d y^2 \}$ and the $\{ e^{a_{\bot}} \}$ is given by the four others. The set $I=2$ is along $y^1, y^3$ and thus partially overlaps the previous one. The set $I=3$ does not, as we take it along $y^4, y^5$. In each set, the various $O_5/D_5$ can be located at different points along their transverse directions, e.g.~some at $y^6=0$ and others at $y^6=\pi$, etc. In \cite{Andriot:2016xvq}, we considered $N=1$, corresponding to only parallel sources; we are interested here in $N> 1$. This excludes the case $p=3$, for which the whole $\mmm$ are the internal transverse directions and there are no internal parallel directions. In other words, in the following, considering $p=3$ forces to take $N=1$. As each set $I$ is defined by its parallel and transverse one-forms, we denote them as $\{ e^{a_{||_I}} \}$ and $\{ e^{a_{\bot_I}} \}$.

Each source action is given by the sum $S_{DBI} + S_{WZ}$, detailed in Appendix \ref{ap:Tmn}. We do not consider higher order corrections to those actions, as e.g.~in \cite{Dasgupta:1999ss}. Only $S_{DBI}$ contributes to the Einstein equation and the dilaton equation of motion (e.o.m.). It does through the energy momentum tensor $T_{MN}$ (here in 10d curved indices) and its trace $T_{10}=g^{MN} T_{MN}$. It is defined as
\beq
\frac{1}{\sqrt{|g_{10}|}} \sum_{{\rm sources}} \frac{\delta S_{DBI}}{\delta g^{MN}} = - \frac{e^{-\phi}}{4 \kappa_{10}^2} T_{M N} \ ,\label{defTmnmain}
\eeq
with the constant $\kappa_{10}$. We will rather use flat indices, $T_{AB}= e^M{}_A e^N{}_B T_{MN}$, and for each source in the sum, we will further decompose onto the different directions with projectors: the 4d flat directions $\alpha$, and the internal $a_{||}$ and $a_{\bot}$. We show in Appendix \ref{ap:Tmn} that for each single source, $T_{a_{\bot}b_{\bot}} = e^M{}_{a_{\bot}} e^N{}_{b_{\bot}} T_{MN}=0$. We then obtain
\beq
T_{AB}= \delta_A^{\alpha} \delta_B^{\beta}\, T_{\alpha \beta} + \sum_I \delta_A^{a_{||_I}} \delta_B^{b_{||_I}} \, T^I_{a_{||_I}b_{||_I}} \ , \label{T1}
\eeq
and explicit expressions for $T_{\alpha \beta}$ and $T^I_{a_{||_I}b_{||_I}}$ are derived in Appendix \ref{ap:Tmn}. The $\sum_{{\rm sources}}$ in \eqref{defTmnmain} gets decomposed into $\sum_{I} \sum_{{\rm sources}\in I}$. Those are present within $T_{\alpha \beta}$, while $T^I_{a_{||_I}b_{||_I}}$ only contains $\sum_{{\rm sources}\in I}$. The trace $T_{10}$ also gets a natural decomposition into traces for each set $I$: $T_{10} = \sum_{I} T_{10}^I$, and expressions for those quantities can be found in Appendix \ref{ap:Tmn}. One eventually shows that
\beq
T_{\alpha \beta}= \eta_{\alpha \beta} \frac{T_{10}}{p+1}  \ , \quad T^I_{a_{||_I}b_{||_I}} = \delta_{a_{||_I}b_{||_I}} \frac{T_{10}^I}{p+1}  \ .\label{T2}
\eeq
Finally, with these definitions, the contribution to the dilaton e.o.m. is given by
\beq
\frac{1}{\sqrt{|g_{10}|}} \sum_{{\rm sources}} \frac{\delta S_{DBI}}{\delta \phi}=- \frac{e^{- \phi}}{2 \kappa_{10}^2} \frac{T_{10}}{p+1} \ .\label{dilcontrib}
\eeq

We now focus on the fluxes Bianchi identities (BI). There is only one BI which includes a source term, because of the single size $p$: the corresponding sourced flux is denoted by the internal form $F_k$ with $ 0 \leq k=8-p \leq 5$. As explained in Appendix \ref{ap:Tmn}, the BI can be written in terms of the previous quantities as
\beq
\d F_k - H \w F_{k-2} = \frac{\varepsilon_p}{p+1}\, \sum_{I} T_{10}^I\, {\rm vol}_{\bot_I}  \ , \label{BI2}
\eeq
with $F_{-1} =F_{-2}=0$ and $\varepsilon_p=(-1)^{p+1} (-1)^{\left[\frac{9-p}{2} \right]}$. We now project the BI on each ${\rm vol}_{\bot_I}$. To that end, we introduce the same notation as in \cite{Andriot:2016xvq}: given a form $G$, the projected form obtained by keeping only its components entirely along directions of ${\rm vol}_{\bot_I}$ is denoted $G|_{\bot_I}$, or $(G)|_{\bot_I}$ if there is an ambiguity. If $G$ is a $(9-p)$-form, the coefficient $(G)_{\bot_I}$ is given by $G|_{\bot_I}= (G)_{\bot_I} {\rm vol}_{\bot_I}$ or equivalently $(G)_{\bot_I} = *_{\bot_I} G|_{\bot_I}$. Projecting the BI \eqref{BI2}, we then get the coefficients
\beq
(\d F_k)_{\bot_I} - (H \w F_{k-2})_{\bot_I} =  \varepsilon_p\, \frac{T_{10}^I}{p+1} \ . \label{BI3}\\
\eeq
Generalizing \cite{Andriot:2016xvq}, one can verify for each $I$ that $(H \w F_{k-2})|_{\bot_I}= H|_{\bot_I} \w F_{k-2}|_{\bot_I}$, and this is also equal to $*_{\bot_I} H|_{\bot_I} \w *_{\bot_I} F_{k-2}|_{\bot_I}= F_{k-2}|_{\bot_I} \w *_{\bot_I}^2 H|_{\bot_I}$. Then, for any sign $\varepsilon$,
\beq
\left|*_{\bot_I} H|_{\bot_I} + \varepsilon e^{\phi}  F_{k-2}|_{\bot_I} \right|^2 = |H|_{\bot_I}|^2 + e^{2\phi} |F_{k-2}|_{\bot_I}|^2 + 2 \varepsilon e^{\phi} (H\w F_{k-2})_{\bot_I} \ , \label{blasquare}
\eeq
where the definition of the square of a form $A$ in $D$ dimensions is $A\w *_{D} A = \d^{D} x\, \sqrt{|g_{D}|}\ |A|^2 $, and here on the $I$-transverse subspace $A|_{\bot_I}\w *_{\bot_I} A|_{\bot_I} = {\rm vol}_{\bot_I}\ |A|_{\bot_I}|^2 $. This will allow us to rewrite the BI.

Finally, the e.o.m. as given in Appendix A of \cite{Andriot:2016xvq} remain valid. We focus here on the dilaton e.o.m. and traces of the Einstein equation. We denote ${\cal R}_{10}= g^{MN} {\cal R}_{MN}$, and
\beq
{\cal R}_4= g^{MN} {\cal R}_{MN=\mu\nu} \ ,\ {\cal R}_6= g^{MN} {\cal R}_{MN=mn}={\cal R}_{10} - {\cal R}_{4} \ , \ (\nabla\del \phi)_4= g^{MN=\mu\nu}\nabla_{M}\del_{N} \phi \ .
\eeq
The dilaton e.o.m., the ten-dimensional Einstein trace, and the four-dimensional one, are
\bea
& \hspace{-0.1in} 2 {\cal R}_{10} + e^{\phi} \frac{T_{10}}{p+1} -|H|^2 + 8(\Delta \phi - |\del \phi|^2 ) = 0 \ ,\label{dileom2}\\
& \hspace{-0.1in} 4 {\cal R}_{10}  + \frac{e^{\phi}}{2} {T}_{10} - |H|^2 - \frac{e^{2\phi}}{2} \sum_{q=0}^6 (5-q) |F_q|^2 -20 |\del \phi|^2 + 18 \Delta \phi = 0\ , \label{10dtrace}\\
& \hspace{-0.1in} {\cal R}_4 - 2{\cal R}_{10} - \frac{2 e^{\phi}}{p+1} {T}_{10} + |H|^2 + e^{2\phi} \sum_{q=0}^6 |F_q|^2  +2 (\nabla\del \phi)_4 + 8 |\del \phi|^2 - 8 \Delta \phi = 0 \ ,\label{4dtrace}
\eea
with only even/odd RR fluxes in IIA/IIB. Note that the above properties on the sources gave $g^{MN} T_{MN=\mu\nu} = 4 T_{10} /(p+1)$. We now have all ingredients needed for the rest of the paper.

\section{Deriving expressions for ${\cal R}_4$}\label{sec:deriv}

We now make use of the tools introduced in Section \ref{sec:formal} to derive interesting expressions and constraints on de Sitter and Minkowski solutions. In this section, we proceed analogously to \cite{Andriot:2016xvq}, generalizing to the case of intersecting sources. As discussed in Section \ref{sec:formal}, we work from now on with a constant dilaton.

\subsection{First derivation and no-go theorem for $p=7,8$}\label{sec:p=7,8}

We first mimic a reasoning made in \cite{Andriot:2015aza, Andriot:2016xvq}, here for a constant dilaton and without warp factor. Using the dilaton e.o.m.~to eliminate ${T}_{10}$ in respectively the ten- and four-dimensional Einstein traces, we get
\bea
&\hspace{-0.1in} (p-3)\left( -2  {\cal R}_{10}  + |H|^2 \right) + 2 |H|^2 - e^{2\phi} \sum_{q=0}^6 (5-q) |F_q|^2 = 0  \label{10dtracesansT10}\\
&\hspace{-0.1in} 3 {\cal R}_4 = - 2{\cal R}_{6} + |H|^2 - e^{2\phi} \sum_{q=0}^6 |F_q|^2 \ .\label{4dtracesansT10}
\eea
Multiplying \eqref{4dtracesansT10} by $(p-3)$ and inserting \eqref{10dtracesansT10}, we obtain
\beq
(p-3) {\cal R}_4  = - 2 |H|^2 + e^{2\phi} \sum_{q=0}^6 (8-q-p) |F_q|^2 \ ,\label{4dtracefinal}
\eeq
or in other words in IIA and IIB
\bea
(p-3) {\cal R}_4  = & - 2 |H|^2 + e^{2\phi} \left( (8-p) |F_0|^2 + (6-p) |F_2|^2 + (4-p) |F_4|^2 + (2-p) |F_6|^2 \right), \label{4dfinalIIA} \\
(p-3) {\cal R}_4  = & - 2 |H|^2 + e^{2\phi} \left( (7-p) |F_1|^2 + (5-p) |F_3|^2 + (3-p) |F_5|^2 \right). \label{4dfinalIIB}
\eea
As explained in \cite{Andriot:2015aza, Andriot:2016xvq}, ${\cal R}_4$ is here only given in terms of the non-sourced fluxes. Indeed, we can rewrite the above as follows, with the notations specified below \eqref{R4T10HF0},
\beq
(p-3) {\cal R}_4  = - 2 |H|^2 + e^{2\phi} (4|F_{k-4}|^2 + 2 |F_{k-2}|^2 -2 |F_{k+2}|^2 - 4 |F_{k+4}|^2 - 6 |F_{k+6}|^2 ) \ ,\label{4dtracefinalFk}
\eeq
and one sees that $F_k$ is absent. We now consider having a de Sitter solution, i.e.~${\cal R}_4>0$: it is clear from above that
\beq
\boxed{\mbox{Result:}}\quad \mbox{There is no de Sitter solution for}\ p=7\ \mbox{or}\ p=8. \label{nogo78}
\eeq
As anticipated in \cite{Andriot:2016xvq}, we prove here that this result holds for intersecting $O_7/D_7$ or $O_8/D_8$ sources, given the few assumptions presented in Section \ref{sec:formal}. Getting de Sitter solutions with intersecting sources of fixed $p$ is then restricted to $p=4,5,6$. To study the latter, we now derive further expressions. The above identities will still appear to be useful.

\subsection{Second derivation}\label{sec:derivgen}

Following and extending \cite{Andriot:2016xvq}, we now combine differently equations of Section \ref{sec:formal}. First, combining the dilaton e.o.m.~and the four-dimensional Einstein trace, we get
\beq
{\cal R}_4 = e^{\phi} \frac{{T}_{10}}{p+1} - e^{2\phi} \sum_{q=0}^6 |F_q|^2 \ , \label{R4T10F}
\eeq
with even/odd RR fluxes in IIA/IIB. We recover the famous requirement for de Sitter solutions, namely that ${T}_{10} > 0$ \cite{Maldacena:2000mw}, i.e.~the need for $O_p$, here in the case of intersecting sources. Combining the dilaton e.o.m.~with the ten-dimensional Einstein trace, we get
\beq
(p-3) e^{\phi} \frac{{T}_{10}}{p+1}  +2 |H|^2 - e^{2\phi} \sum_{q=0}^6 (5-q) |F_q|^2 = 0\ . \label{10dtracesansR10}
\eeq
Equation \eqref{R4T10F} is multiplied by $-(p+1)$, and added to \eqref{10dtracesansR10}, to give
\beq
{\cal R}_4 = -\frac{1}{p+1} \bigg( -4 e^{\phi} \frac{{T}_{10}}{p+1} +2|H|^2 + e^{2\phi} \sum_{q=0}^6 (p+q-4) |F_q|^2  \bigg) \ . \label{R4T10HF0}
\eeq
From now on, we use notations of \eqref{BI2}, where the magnetically sourced flux is $F_k$ with $0 \leq k=8-p \leq 5$, and $F_{-1}=F_{-2}=F_7=F_8=F_9=F_{10}=F_{11}=0$. Then, \eqref{R4T10HF0} gets rewritten as
\beq
{\cal R}_4 = -\frac{2}{p+1} \bigg( -2 e^{\phi} \frac{{T}_{10}}{p+1} +|H|^2  + e^{2\phi} (  |F_{k-2}|^2 + 2 |F_{k}|^2 + 3 |F_{k+2}|^2 + 4 |F_{k+4}|^2 + 5 |F_{k+6}|^2 )  \bigg)  \ . \nn
\eeq
We now replace $T_{10}= \sum_I T_{10}^I$ using the projected BI \eqref{BI3}. The sum on $I$ offers several ways to proceed. We choose one here and discuss other possibilities in Section \ref{sec:susymink} and Appendix \ref{ap:HF}. Using \eqref{blasquare}, we get
\bea
{\cal R}_4  = -\frac{2}{p+1} \bigg( & -  2 \varepsilon_p e^{\phi} \sum_I (\d F_k)_{\bot_I}  + \sum_I \left|*_{\bot_I}H|_{\bot_I} + \varepsilon_p e^{\phi} F_{k-2}|_{\bot_I} \right|^2  \label{firstsquaregen}\\
& +|H|^2 - \sum_I |H|_{\bot_I}|^2 + e^{2\phi} (  |F_{k-2}|^2 - \sum_I |F_{k-2}|_{\bot_I}|^2 ) \nn\\
& + e^{2\phi} ( 2 |F_{k}|^2 + 3 |F_{k+2}|^2 + 4 |F_{k+4}|^2 + 5 |F_{k+6}|^2 )  \bigg)  \ . \nn
\eea
At this stage with $N=1$, i.e.~parallel sources  \cite{Andriot:2016xvq}, we could already obtain a no-go theorem. For $N>1$, we cannot be as conclusive due to $|H|^2 - \sum_I |H|_{\bot_I}|^2$ and $ |F_{k-2}|^2 - \sum_I |F_{k-2}|_{\bot_I}|^2$, whose signs are not necessarily positive, especially with an overlap of transverse directions. Consider the following example: $N=2$ sets of sources with $p=5$, along internal $e^1\w e^2$ and $e^1\w e^3$, and $H=h\, e^4\w e^5 \w e^6$. One has $H|_{\bot_1}= h\, e^4\w e^5 \w e^6 = H|_{\bot_2}$, there is an overlap of components in transverse directions.\footnote{Note that $H|_{\bot_I}$ is typically non-zero, e.g.~in the simple case of constant components, since the $H$-flux is odd under the orientifold projection, $\sigma(H) = - H$ \cite{Koerber:2007hd}.} One deduces $\sum_I |H|_{\bot_I}|^2 = 2 h^2 \geq h^2 = |H|^2$. Since $H$ has three indices, one may find other situations where $|H|^2 - \sum_I |H|_{\bot_I}|^2 \geq 0$, e.g.~in the case where $p=5$ sources do not overlap. But it is more difficult for $F_{k-2}$ which has less indices.

To proceed further, we rewrite for each $I$ the term $(\d F_k)_{\bot_I}$. We recall the definition of the quantity $f^{a}{}_{bc}$, not necessarily constant
\beq
\d e^a= -\frac{1}{2} f^{a}{}_{bc} e^b\w e^c \ \Leftrightarrow \ f^{a}{}_{bc} = 2 e^a{}_m \del_{[b} e^m{}_{c]} = - 2 e^m{}_{[c} \del_{b]} e^a{}_{m} \label{fabc} \ .
\eeq
For each $I$, we can choose to decompose the flux $F_k$ (a priori independent of $I$) on the corresponding parallel or transverse flat components
\beq
F_k= \frac{1}{k!} F^{(0)_I}_{k\ a_{1\bot_I}\dots a_{k\bot_I}} e^{a_{1\bot_I}} \w \dots \w e^{a_{k\bot_I}} + \frac{1}{(k-1)!} F^{(1)_I}_{k\ a_{1||_I}\dots a_{k\bot_I}} e^{a_{1||_I}} \w e^{a_{2\bot_I}} \w \dots \w e^{a_{k\bot_I}} + \dots \nn
\eeq
By definition, $F_k^{(0)_I}=F_k|_{\bot_I}$; we also choose the convenient notation $F_0|_{\bot_I}=F_0$ and $F_0^{(1)_I}=0$. As a consequence, one gets
\beq
\forall I,\ |F_k|^2 = \sum_{n=0}^{p-3} |F_k^{(n)_I}|^2 \ .\label{Fksquare}
\eeq
That sum may end before $p-3$, depending on $k=8-p$. One deduces
\beq
(\d F_k)|_{\bot_I}  = (\d F_k^{(0)_I})|_{\bot_I} + (\d F_k^{(1)_I})|_{\bot_I}\ , \ \ (\d F_k^{(1)_I})|_{\bot_I} =  \iota_{a_{||_I}} F_k^{(1)_I} \w (\d e^{a_{||_I}})|_{\bot_I} \ ,\label{dFgeneral}
\eeq
where $(\d e^{a_{||_I}})|_{\bot_I} = -\frac{1}{2} f^{a_{||_I}}{}_{b_{\bot_I}c_{\bot_I}} e^{b_{\bot_I}}\w e^{c_{\bot_I}}$, and the contraction by a vector $\del_{a_{||_I}}$ is given by $\iota_{a_{||_I}} e^{b_{||_I}} = \delta_{a_{||_I}}^{b_{||_I}}$. Similarly to \eqref{blasquare}, we get
\bea
& \sum_{a_{||_I}} \left| *_{\bot_I}( \d e^{a_{||_I}})|_{\bot_I} - \varepsilon_p e^{\phi}\, \iota_{a_{||_I}} F_k^{(1)_I}  \right|^2 = \sum_{a_{||_I}} e^{2\phi} |\iota_{a_{||_I}} F_k^{(1)_I}|^2 + \sum_{a_{||_I}} |(\d e^{a_{||_I}})|_{\bot_I}|^2 \label{squareF1}\\
& \phantom{\sum_{a_{||_I}} \left| *_{\bot}( \d e^{a_{||_I}})|_{\bot_I} - \varepsilon_p e^{\phi}\, \iota_{a_{||_I}} F_k^{(1)_I}  \right|^2 =} -  2 \varepsilon_p e^{\phi} (\iota_{a_{||_I}} F_k^{(1)_I} \w (\d e^{a_{||_I}})|_{\bot_I})_{\bot_I} \nn\\
& \mbox{with}\ \sum_{a_{||_I}} e^{2\phi} |\iota_{a_{||_I}} F_k^{(1)_I}|^2 = e^{2\phi} |F_k^{(1)_I}|^2 \ ,\nn\\
& \mbox{and}\ \sum_{a_{||_I}} |(\d e^{a_{||_I}})|_{\bot_I}|^2 = \frac{1}{2} \delta^{be} \delta^{cf} \delta_{ad} f^{a_{||_I}}{}_{b_{\bot_I}c_{\bot_I}} f^{d_{||_I}}{}_{e_{\bot_I}f_{\bot_I}} \ .\label{de2}
\eea
For each $I$, one can thus rewrite
\bea
-  2 \varepsilon_p e^{\phi} (\d F_k)_{\bot_I} = & -  2 \varepsilon_p e^{\phi} (\d F_k^{(0)_I})_{\bot_I} + \sum_{a_{||_I}} \left| *_{\bot_I}( \d e^{a_{||_I}})|_{\bot_I} - \varepsilon_p e^{\phi}\, \iota_{a_{||_I}} F_k^{(1)_I} \right|^2 \label{dFrewrite} \\
& \phantom{-  2 \varepsilon_p e^{\phi} (\d F_k^{(0)_I})_{\bot_I}}\ - e^{2\phi} |F_k^{(1)_I}|^2 - \sum_{a_{||_I}} |(\d e^{a_{||_I}})|_{\bot_I}|^2  \ . \nn
\eea

To accommodate the very last term, rewritten in \eqref{de2}, we now need to introduce part of the internal curvature. We consider the trace of the Einstein equation along internal parallel directions for one given $J$. We denote ${\cal R}_{6||_J}= \eta^{AB}{\cal R}_{AB=a_{||_J}b_{||_J}}$. We obtain the same result as in \cite{Andriot:2016xvq} (using the four-dimensional Einstein trace \eqref{4dtrace}) up to new $T_{10}^I$ contributions
\bea
{\cal R}_{6||_J} &= \frac{p-3}{4} \left({\cal R}_4 +  2e^{2\phi} |F_{6}|^2  \right) + \frac{e^{\phi}}{2} \left( T_{a_{||_J}}^{a_{||_J}} - \frac{p-3}{p+1} T_{10} \right) \label{traceparIIA}\\
& + \frac{1}{2} \left(|H|^2 - |H|_{\bot_J}|^2 + e^{2\phi} (  |F_{2}|^2 - |F_{2}|_{\bot_J}|^2 + |F_{4}|^2 - |F_{4}|_{\bot_J}|^2 \right) \nn \\
& + \frac{1}{2} \sum_{n=2}^{p-3} (n-1) \left(|H^{(n)_J}|^2 + e^{2\phi} (  |F_2^{(n)_J}|^2 + |F_4^{(n)_J}|^2 ) \right) \nn\\
{\cal R}_{6||_J} &= \frac{p-3}{4} \left({\cal R}_4 + e^{2\phi} |F_{5}|^2  \right) + \frac{e^{\phi}}{2} \left( T_{a_{||_J}}^{a_{||_J}} - \frac{p-3}{p+1} T_{10} \right) \label{traceparIIB}\\
& + \frac{1}{2} \left(|H|^2 - |H|_{\bot_J}|^2 + e^{2\phi} (  |F_{1}|^2 - |F_{1}|_{\bot_J}|^2 + |F_{3}|^2 - |F_{3}|_{\bot_J}|^2 \right) \nn\\
& + \frac{1}{4} e^{2\phi} \left(  |F_{5}|^2 - |F_{5}|_{\bot_J}|^2 - |*_6 F_{5}|^2 + |(*_6 F_{5})|_{\bot_J}|^2 \right) \nn\\
& + \frac{1}{2} \sum_{n=2}^{p-3} (n-1) \left(|H^{(n)_J}|^2 + e^{2\phi} (  |F_3^{(n)_J}|^2 + \frac{1}{2} |F_5^{(n)_J}|^2 ) \right)  \ ,\nn
\eea
where we denote $T_{a_{||_J}}^{a_{||_J}}= \eta^{AB}T_{AB=a_{||_J}b_{||_J}}$. This is computed, thanks to \eqref{T1} and \eqref{T2}, to be
\beq
T_{a_{||_J}}^{a_{||_J}} = \frac{p-3}{p+1} T_{10}^J + \sum_{I\neq J} \frac{\delta_{a_{||_J}}^{a_{||_I}}}{p+1} T_{10}^I  \ ,\label{newT}
\eeq
where $\delta_{a_{||_J}}^{a_{||_I}}$ counts the number of common parallel internal directions between the sets $I$ and $J$, non-zero if there is an overlap, and smaller than $p-3$ by definition. The parallel directions of the sets $I$ and $J$ are not necessarily defined in the same orthonormal basis, so the number $\delta_{a_{||_J}}^{a_{||_I}}$ may actually not be an integer; the notation is then formal. We rewrite \eqref{traceparIIA} and \eqref{traceparIIB} as follows for $0 \leq k=8-p \leq 5$ (for $p=3$, all internal directions are transverse so any term with internal parallel direction is taken to vanish)
\bea
\hspace{-0.3in} 2 {\cal R}_{6||_J} &  -  \frac{p-3}{2}  {\cal R}_4 - e^{\phi} \left( T_{a_{||_J}}^{a_{||_J}}  - \frac{p-3}{p+1} T_{10} \right) = |H|^2 - |H|_{\bot_J}|^2 + e^{2\phi} \left(  |F_{k-2}|^2 - |F_{k-2}|_{\bot_J}|^2 \right) \nn\\
& \!\!\!\! + e^{2\phi} \bigg(  |F_{k}|^2 - |F_{k}|_{\bot_J}|^2 + |F_{k+2}|^2 + (9-p) |F_{k+4}|^2 + 5 |F_{k+6}|^2 + \frac{1}{2} ( |(*_6 F_{5})|_{\bot_J}|^2 - |F_{5}|_{\bot_J}|^2 ) \bigg) \nn\\
& \!\!\!\! + \sum_{n=2}^{p-3} (n-1) \left(|H^{(n)_J}|^2 + e^{2\phi} (  |F_k^{(n)_J}|^2 + |F_{k+2}^{(n)_J}|^2 + \frac{p-6}{2} |F_{k+4}^{(n)_J}|^2 + \frac{p-7}{4} |F_5^{(n)_J}|^2 ) \right) \label{tracepargen}
\eea
where the $F_{5}$ terms should only be considered in IIB. In addition, for each $J$, ${\cal R}_{6||_J}$ can be computed as in \cite{Andriot:2016xvq}: one obtains, without warp factor,
\beq
{\cal R}_{6||_J} =  {\cal R}_{||_J} +  {\cal R}_{||_J}^{\bot_J} + \frac{1}{2} \sum_{a_{||_J}} |(\d e^{a_{||_J}})|_{\bot_J}|^2 \ , \label{R6par}
\eeq
with the following curvature terms (we drop for simplicity the label $J$ on each $a_{||_J}$ and $a_{\bot_J}$)
\bea
2 {\cal R}_{||}  =&\ 2 \delta^{cd} \del_{c_{||}} f^{a_{||}}{}_{d_{||}a_{||}} -\delta^{ab} f^{d_{||}}{}_{c_{||}a_{||}} f^{c_{||}}{}_{d_{||}b_{||}}- \frac{1}{2} \delta^{ch}\delta^{dj}\delta_{ab} f^{a_{||}}{}_{c_{||}j_{||}} f^{b_{||}}{}_{h_{||}d_{||}} \ , \label{Rpar}\\
2 {\cal R}_{||}^{\bot}  =&  -\delta^{ab} f^{d_{\bot}}{}_{c_{\bot}a_{||}} f^{c_{\bot}}{}_{d_{\bot}b_{||}} - \delta^{ab} \delta^{dg} \delta_{ch} f^{h_{\bot}}{}_{g_{\bot}a_{||}} f^{c_{\bot}}{}_{d_{\bot}b_{||}} \label{Rbotpar}\\
& - 2 \delta^{ab} f^{d_{\bot}}{}_{c_{||}a_{||}} f^{c_{||}}{}_{d_{\bot}b_{||}} - \delta^{ab} \delta^{dg} \delta_{ch} f^{h_{\bot}}{}_{g_{||}a_{||}} f^{c_{\bot}}{}_{d_{||}b_{||}} \ .\nn
\eea

We can now derive the general ${\cal R}_4$ expression, analogous to that of \cite{Andriot:2016xvq}: combining \eqref{dFrewrite}, \eqref{tracepargen} and \eqref{R6par}, we get
\bea
& -  2 \varepsilon_p e^{\phi} (\d F_k)_{\bot_I} + |H|^2 - |H|_{\bot_I}|^2 + e^{2\phi} \left(  |F_{k-2}|^2 - |F_{k-2}|_{\bot_I}|^2 \right) \\
= & -  2 \varepsilon_p e^{\phi} (\d F_k^{(0)_I})_{\bot_I} + \sum_{a_{||_I}} \left| *_{\bot_I}( \d e^{a_{||_I}})|_{\bot_I} - \varepsilon_p e^{\phi} \, \iota_{a_{||_I}} F_k^{(1)_I} \right|^2 - e^{2\phi} |F_k^{(1)_I}|^2 \nn\\
& + 2{\cal R}_{||_I} +  2 {\cal R}_{||_I}^{\bot_I} -  \frac{p-3}{2}  {\cal R}_4 - e^{\phi} \left( T_{a_{||_I}}^{a_{||_I}}  - \frac{p-3}{p+1} T_{10} \right) \nn\\
& - e^{2\phi} \bigg(  |F_{k}|^2 - |F_{k}|_{\bot_I}|^2 + |F_{k+2}|^2 + (9-p) |F_{k+4}|^2 + 5 |F_{k+6}|^2 + \frac{1}{2} ( |(*_6 F_{5})|_{\bot_I}|^2 - |F_{5}|_{\bot_I}|^2 ) \bigg) \nn\\
& - \sum_{n=2}^{p-3} (n-1) \left(|H^{(n)_I}|^2 + e^{2\phi} (  |F_k^{(n)_I}|^2 + |F_{k+2}^{(n)_I}|^2 + \frac{p-6}{2} |F_{k+4}^{(n)_I}|^2 + \frac{p-7}{4} |F_5^{(n)_I}|^2 ) \right) \nn \ .
\eea
Replacing in \eqref{firstsquaregen}, this gives
\bea
 \frac{(1-N)p+3N+1}{2} {\cal R}_4  =\ &   2 \varepsilon_p e^{\phi} \sum_I (\d F_k^{(0)_I})_{\bot_I} - \sum_I  2 e^{2\phi} |F_k^{(0)_I}|^2  - \sum_I \left|*_{\bot_I}H|_{\bot_I} + \varepsilon_p e^{\phi} F_{k-2}|_{\bot_I} \right|^2 \nn \\
& - \sum_I \sum_{a_{||_I}} \left| *_{\bot_I}( \d e^{a_{||_I}})|_{\bot_I} - \varepsilon_p e^{\phi}\, \iota_{a_{||_I}} F_k^{(1)_I} \right|^2 \\
& - \sum_I 2( {\cal R}_{||_I} +  {\cal R}_{||_I}^{\bot_I} )  + \sum_I e^{\phi} \left( T_{a_{||_I}}^{a_{||_I}}  - \frac{p-3}{p+1} T_{10} \right) \nn\\
& + (N-1) \left( |H|^2 +  e^{2\phi}   |F_{k-2}|^2 +  e^{2\phi}  2 |F_{k}|^2 \right) \nn\\
& - \sum_I e^{2\phi} (  |F_{k}|^2 -|F_k^{(0)_I}|^2 - |F_k^{(1)_I}|^2) \nn\\
& + e^{2\phi} \bigg( (N-3) |F_{k+2}|^2 + (N(9-p)-4) |F_{k+4}|^2 + 5(N-1) |F_{k+6}|^2 \bigg)\nn\\
& + \sum_I  \frac{1}{2} e^{2\phi} \bigg( |(*_6 F_{5})|_{\bot_I}|^2 - |F_{5}|_{\bot_I}|^2 \bigg) \nn\\
& \hspace{-1.2in} + \sum_I \sum_{n=2}^{p-3} (n-1) \left(|H^{(n)_I}|^2 + e^{2\phi} (  |F_k^{(n)_I}|^2 + |F_{k+2}^{(n)_I}|^2 + \frac{p-6}{2} |F_{k+4}^{(n)_I}|^2 + \frac{p-7}{4} |F_5^{(n)_I}|^2 ) \right) \ . \nn
\eea
This generalizes the expression obtained in \cite{Andriot:2016xvq}. There are two new types of terms, vanishing for parallel sources: those in $T_{10}^I$, and those with fluxes times $N-1$. We now rewrite these new terms using combinations of previous equations. First, using \eqref{newT} and $T_{10}=\sum_I T^I_{10}$, we deduce
\beq
\sum_I e^{\phi} \left( T_{a_{||_I}}^{a_{||_I}}  - \frac{p-3}{p+1} T_{10} \right) = e^{\phi} (1 - N) \frac{p-3}{p+1} T_{10} + e^{\phi} \sum_I \sum_{J\neq I} \frac{\delta_{a_{||_I}}^{a_{||_J}}}{p+1} T_{10}^J  \ . \label{Tterms}
\eeq
We further replace $T_{10}$ in the first term using \eqref{R4T10F}. The above is then rewritten as
\bea
\frac{N(p-3) +7-p}{2} {\cal R}_4  =\ &   2 \varepsilon_p e^{\phi} \sum_I (\d F_k^{(0)_I})_{\bot_I} - \sum_I  2 e^{2\phi} |F_k^{(0)_I}|^2  - \sum_I \left|*_{\bot_I}H|_{\bot_I} + \varepsilon_p e^{\phi} F_{k-2}|_{\bot_I} \right|^2 \nn \\
& - \sum_I \sum_{a_{||_I}} \left| *_{\bot_I}( \d e^{a_{||_I}})|_{\bot_I} - \varepsilon_p e^{\phi}\, \iota_{a_{||_I}} F_k^{(1)_I} \right|^2 \label{R4bla}\\
& - \sum_I 2( {\cal R}_{||_I} +  {\cal R}_{||_I}^{\bot_I} )  + \sum_I (|H^{(2)_I}|^2 + 2 |H^{(3)_I}|^2 ) + e^{\phi} \sum_I \sum_{J\neq I} \frac{\delta_{a_{||_I}}^{a_{||_J}}}{p+1} T_{10}^J  \nn\\
& + (N-1) \left( |H|^2 +  e^{2\phi} ( (4-p) |F_{k-2}|^2 +  (5-p) |F_{k}|^2 ) \right) \nn\\
& - e^{2\phi} \sum_I  (  |F_{k}|^2 -|F_k^{(0)_I}|^2 - |F_k^{(1)_I}|^2) - e^{2\phi} (N-1)(p-3) |F_{k-4}|^2 \nn\\
& + e^{2\phi} \bigg( ((N-1)(4-p) -2) |F_{k+2}|^2 + (2(N-1)(6-p) + (5-p)) |F_{k+4}|^2 \bigg)\nn\\
& + e^{2\phi} \sum_I  \frac{1}{2}  \bigg( |(*_6 F_{5})|_{\bot_I}|^2 - |F_{5}|_{\bot_I}|^2 \bigg) \nn\\
& + e^{2\phi} \sum_I \sum_{n=2}^{p-3} (n-1) \left(  |F_k^{(n)_I}|^2 + |F_{k+2}^{(n)_I}|^2 + \frac{p-6}{2} |F_{k+4}^{(n)_I}|^2 + \frac{p-7}{4} |F_5^{(n)_I}|^2  \right) \ . \nn
\eea
Secondly, considering the flux terms proportional to $N-1$, we use \eqref{4dtracefinalFk} to replace $|H|^2$. We rewrite the above as
\bea
\hspace{-0.3in} ((N-1)(p-3) + 2) {\cal R}_4  =\  &   2 \varepsilon_p e^{\phi} \sum_I (\d F_k^{(0)_I})_{\bot_I} - \sum_I  2 e^{2\phi} |F_k^{(0)_I}|^2  - \sum_I \left|*_{\bot_I}H|_{\bot_I} + \varepsilon_p e^{\phi} F_{k-2}|_{\bot_I} \right|^2 \nn \\
& - \sum_I \sum_{a_{||_I}} \left| *_{\bot_I}( \d e^{a_{||_I}})|_{\bot_I} - \varepsilon_p e^{\phi}\, \iota_{a_{||_I}} F_k^{(1)_I}  \right|^2 \label{final}\\
& - \sum_I 2( {\cal R}_{||_I} +  {\cal R}_{||_I}^{\bot_I} ) + \sum_I (|H^{(2)_I}|^2 + 2 |H^{(3)_I}|^2 )  + e^{\phi} \sum_I \sum_{J\neq I} \frac{\delta_{a_{||_I}}^{a_{||_J}}}{p+1} T_{10}^J  \nn\\
& - (N-1) (p-5) e^{2\phi}( |F_{k-4}|^2 + |F_{k-2}|^2 + |F_{k}|^2 ) \nn\\
& - e^{2\phi} \sum_I  (  |F_{k}|^2 -|F_k^{(0)_I}|^2 - |F_k^{(1)_I}|^2) + e^{2\phi} \sum_I  \frac{1}{2}  \big( |(*_6 F_{5})|_{\bot_I}|^2 - |F_{5}|_{\bot_I}|^2 \big) \nn\\
& \hspace{-0.7in} + e^{2\phi} \bigg( ((N-1)(3-p) -2) |F_{k+2}|^2 + (2N-1)(5-p) |F_{k+4}|^2 -3(N-1) |F_{k+6}|^2 \bigg)\nn\\
& \hspace{-0.7in} + e^{2\phi} \sum_I \sum_{n=2}^{p-3} (n-1) \left(  |F_k^{(n)_I}|^2 + |F_{k+2}^{(n)_I}|^2 + \frac{p-6}{2} |F_{k+4}^{(n)_I}|^2 + \frac{p-7}{4} |F_5^{(n)_I}|^2  \right) \ . \nn
\eea
The coefficient of ${\cal R}_4$ on the left-hand side is always positive. The last three lines of fluxes are always negative, so we denote them $- e^{2\phi} ( \mbox{fluxes} )$ in the following: they are given by
\bea
& p=3:\ - e^{2\phi} ( \mbox{fluxes} ) = 0 \\
& p=4:\ - e^{2\phi} ( \mbox{fluxes} ) = - (N+1) e^{2\phi} |F_{6}|^2 \nn\\
& p=5:\ - e^{2\phi} ( \mbox{fluxes} ) = - e^{2\phi} \sum_I (2|F_{5}|^2 - \frac{1}{2} |(*_6 F_{5})|_{\bot_I}|^2 - \frac{1}{2} |F_5^{(2)_I}|^2 ) \nn\\
& p=6:\ - e^{2\phi} ( \mbox{fluxes} ) = - e^{2\phi} \Big( (3N-1) |F_{4}|^2 -\sum_I ( |F_4^{(2)_I}|^2 + 2 |F_4^{(3)_I}|^2 )  + (2N-1) |F_{6}|^2 \Big) \nn\\
& p=7:\ - e^{2\phi} ( \mbox{fluxes} ) = - e^{2\phi} \Big( 2(2N-1) |F_{3}|^2 -\sum_I ( |F_3^{(2)_I}|^2 + 2 |F_3^{(3)_I}|^2 ) \Big)  -2  e^{2\phi} (N-1) |F_{5}|^2 \nn\\
& \phantom{p=7:\ - e^{2\phi} ( \mbox{fluxes} ) = } -2  e^{2\phi} \sum_I \big( |F_{5}|^2 - \frac{1}{4} |(*_6 F_{5})|_{\bot_I}|^2 - \sum_{n=2}^4 \frac{n-1}{4} |F_5^{(n)_I}|^2 \big)  \nn\\
& p=8:\ - e^{2\phi} ( \mbox{fluxes} ) = - e^{2\phi} \Big( (5N-3) |F_{2}|^2 -\sum_I |F_2^{(2)_I}|^2 + 3(2N-1) |F_{4}|^2 -\sum_I \sum_{n=2}^4 (n-1) |F_4^{(n)_I}|^2 \nn\\
& \phantom{p=8:\ - e^{2\phi} ( \mbox{fluxes} ) = - e^{2\phi} \Big(}  + 3(N-1) |F_{6}|^2 \Big) \ .\nn
\eea
To verify that these lines are negative, we use \eqref{Fksquare} and that $|F_{5}|^2 = |*_6 F_{5}|^2 \geq |(*_6 F_{5})|_{\bot_I}|^2$. Interestingly, all $F_{k}$ terms have been canceled. We thus rewrite \eqref{final} as follows
\bea
\hspace{-0.3in} ((N-1)(p-3) + 2) {\cal R}_4  =\  &   2 \varepsilon_p e^{\phi} \sum_I (\d F_k^{(0)_I})_{\bot_I} - \sum_I  2 e^{2\phi} |F_k^{(0)_I}|^2  - \sum_I \left|*_{\bot_I}H|_{\bot_I} + \varepsilon_p e^{\phi} F_{k-2}|_{\bot_I} \right|^2 \nn \\
& - \sum_I \sum_{a_{||_I}} \left| *_{\bot_I}( \d e^{a_{||_I}})|_{\bot_I} - \varepsilon_p e^{\phi}\, \iota_{a_{||_I}} F_k^{(1)_I}  \right|^2 \label{finalbis}\\
& + \sum_I (-2 {\cal R}_{||_I} -2  {\cal R}_{||_I}^{\bot_I} + |H^{(2)_I}|^2 + 2 |H^{(3)_I}|^2 )  + e^{\phi} \sum_I \sum_{J\neq I} \frac{\delta_{a_{||_I}}^{a_{||_J}}}{p+1} T_{10}^J  \nn\\
& - (N-1) (p-5) e^{2\phi}( |F_{k-4}|^2 + |F_{k-2}|^2 + |F_{k}|^2 ) \nn\\
& \hspace{-1.3in} - e^{2\phi} \bigg( ((N-1)(p-3) + 2) |F_{k+2}|^2 + (2N-1)(p-5) |F_{k+4}|^2 + 3(N-1) |F_{k+6}|^2 \bigg)\nn\\
& \hspace{-1.3in} + e^{2\phi} \sum_I \bigg( \frac{1}{2}  \big( |(*_6 F_{5})|_{\bot_I}|^2 - |F_{5}|_{\bot_I}|^2 \big) + \sum_{n=2}^{p-3} (n-1) \big( |F_{k+2}^{(n)_I}|^2 + \frac{p-6}{2} |F_{k+4}^{(n)_I}|^2 + \frac{p-7}{4} |F_5^{(n)_I}|^2  \big) \bigg) \ , \nn
\eea
where the last two lines are the above negative combinations $- e^{2\phi} ( \mbox{fluxes} )$. Expression \eqref{finalbis} and its various signs are the starting point for the subsequent analysis.

\section{No-go theorems for $p=4,5,6$}\label{sec:nogo456}

The expressions derived in Section \ref{sec:deriv} are valid for $3 \leq p \leq 8$, and we now use them to get conditions on de Sitter solutions. As obtained in Section \ref{sec:p=7,8}, de Sitter solutions are however not possible in our setting for $p=3,7,8$, so we focus on the three remaining cases.

\subsection{First considerations and the $p=4$ case}\label{sec:p=4}

We start by integrating \eqref{finalbis} over the internal manifold. As a generalization of \cite{Andriot:2016xvq}, $(\d F_k^{(0)_I})_{\bot_I}$ gets integrated to zero for each $I$, using the requirement \eqref{trace} and that the compact internal manifold has no boundary. Indeed, we proceed as follows for each $I$
\bea
\int_6 {\rm vol}_6 (\d F_k^{(0)_I})_{\bot_I} &=  \int_6 {\rm vol}_{||_I} \w  (\d F_k^{(0)_I})|_{\bot_I} =  \int_6 {\rm vol}_{||_I} \w  \d F_k^{(0)_I} = (-1)^p \int_6 \d {\rm vol}_{||_I} \w   F_k^{(0)_I} \nn\\
& = (-1)^{p+1} \int_6 f^{a_{||_I}}{}_{b_{\bot_I}a_{||_I}} e^{b_{\bot_I}} \w {\rm vol}_{||_I} \w   F_k^{(0)_I} \ , \label{inttrick}
\eea
and $f^{a_{||_I}}{}_{b_{\bot_I}a_{||_I}} =- f^{a_{\bot_I}}{}_{b_{\bot_I}a_{\bot_I}} = 0 $. We deduce
\bea
\boxed{\mbox{Result:}}\quad ((N-1)(p-3) & + 2 ) \, {\cal R}_4 \int_6 {\rm vol}_6 \label{finalint}\\
=  - \int_6 {\rm vol}_6 \bigg( & \sum_I  2 e^{2\phi} |F_k^{(0)_I}|^2  + \sum_I \left|*_{\bot_I}H|_{\bot_I} + \varepsilon_p e^{\phi} F_{k-2}|_{\bot_I} \right|^2 \nn \\
& + \sum_I \sum_{a_{||_I}} \left| *_{\bot_I}( \d e^{a_{||_I}})|_{\bot_I} - \varepsilon_p e^{\phi}\, \iota_{a_{||_I}} F_k^{(1)_I} \right|^2  + e^{2\phi} ( \mbox{fluxes} )  \nn\\
& + (N-1) (p-5) e^{2\phi}( |F_{k-4}|^2 + |F_{k-2}|^2 + |F_{k}|^2 ) \nn \\
& + \sum_I ( 2{\cal R}_{||_I} + 2 {\cal R}_{||_I}^{\bot_I} - |H^{(2)_I}|^2 - 2 |H^{(3)_I}|^2 )  - e^{\phi} \sum_I \sum_{J\neq I} \frac{\delta_{a_{||_I}}^{a_{||_J}}}{p+1} T_{10}^J \bigg) \ ,\nn
\eea
where we recall that the terms $( \mbox{fluxes} )$ are positive.

We now focus on the last two lines of \eqref{finalint}, since their sign is a priori not settled. To start with, one recovers the combination of curvature terms and $H$-flux components, as a generalization of the case of parallel sources \cite{Andriot:2016xvq}. The two other terms are new and vanish for $N=1$. The flux terms proportional to $(N-1)(p-5)$ indicate a surprising distinction to be made between $p=4$ and the higher values. This distinction is present as well through the last term in $\delta_{a_{||_I}}^{a_{||_J}}$. Indeed, this term is only non-zero if there is an overlap of the sources (see below \eqref{newT}). However for $p=4$, the sources in two sets $I \neq J$ cannot overlap because they only have one internal direction and should not be parallel.\footnote{This might be refined by considering sources at angles smaller than $\tfrac{\pi}{2}$, even though one may also consider in that case a projection on an orthogonal basis.} So this last term vanishes for $p=4$. The same holds for $|H^{(2)_I}|^2, |H^{(3)_I}|^2$. We deduce the following requirement for a de Sitter solution, which could also be turned into a no-go theorem
\bea
& \mbox{For $p=4$, having a de Sitter solution requires} \label{requirep=4}\\
& \int_6 {\rm vol}_6 \bigg( \sum_I 2( {\cal R}_{||_I} +  {\cal R}_{||_I}^{\bot_I} ) - (N-1) e^{2\phi}( |F_{0}|^2 + |F_{2}|^2 + |F_{4}|^2 ) + (N+1) e^{2\phi} |F_{6}|^2 \bigg) < 0 \ .\nn
\eea
This seems easy to achieve, so the constraints on the $p=4$ case are unexpectedly loose. For $p\geq 5$, we obtain similarly from \eqref{finalint} different requirements, that we summarize in \eqref{require3}.

We turn to constraints deduced from the internal trace on parallel directions, for $p>3$. The right-hand side of \eqref{tracepargen} is positive. We deduce that a de Sitter solution requires
\beq
 2 {\cal R}_{6||_J}   - |H^{(2)_J}|^2 - 2 |H^{(3)_J}|^2   - e^{\phi} \left( T_{a_{||_J}}^{a_{||_J}}  - \frac{p-3}{p+1} T_{10} \right) > 0 \ .
\eeq
While one could infer more conditions by developing the $T_{10}$ terms, we rather sum over $J$ and use \eqref{R6par}, \eqref{Tterms}, to get the following requirement for $p>3$
\bea
\sum_I (2 {\cal R}_{||_I} + 2 {\cal R}_{||_I}^{\bot_I}  - |H^{(2)_I}|^2 & - 2 |H^{(3)_I}|^2) - e^{\phi} \sum_I \sum_{J\neq I} \frac{\delta_{a_{||_I}}^{a_{||_J}}}{p+1} T_{10}^J  \label{require2} \\
& + \sum_I \sum_{a_{||_I}} |(\d e^{a_{||_I}})|_{\bot_I}|^2 + e^{\phi} (N-1) \frac{p-3}{p+1} T_{10}  > 0 \ . \nn
\eea

Combined with the requirement obtained from \eqref{finalint} for $p\geq 5$, we deduce
\bea
\boxed{\mbox{Result:}}\quad & \hspace{0.5in} \mbox{For $p\geq 5$, having a de Sitter solution requires} \label{require3}\\
& \int_6 {\rm vol}_6 \bigg( - \sum_I \sum_{a_{||_I}} |(\d e^{a_{||_I}})|_{\bot_I}|^2 - e^{\phi} (N-1) \frac{p-3}{p+1} T_{10} \bigg) \nn\\
&  < \int_6 {\rm vol}_6 \bigg( \sum_I ( 2 {\cal R}_{||_I} + 2 {\cal R}_{||_I}^{\bot_I}  - |H^{(2)_I}|^2 - 2 |H^{(3)_I}|^2 )  - e^{\phi} \sum_I \sum_{J\neq I} \frac{\delta_{a_{||_I}}^{a_{||_J}}}{p+1} T_{10}^J \bigg) < 0 \ ,\nn
\eea
which can as well be turned into no-go theorems. Let us compare this formula to the one obtained in the case of parallel sources \cite{Andriot:2016xvq}. A first difference is the presence of sums over the sets $I$. This makes the contributions of curvature terms more likely to be non-vanishing. Indeed, they tend to be all negative (see Section \ref{sec:groupmanif}), so having one of them non-zero would be enough. The presence of the two source terms are two other differences. The contribution in $(N-1) T_{10}$ lowers the bound on the left-hand side.\footnote{\label{foot:signT10I}The inequality between the two source terms is natural, although not systematic, as we briefly explain here. The sign of each $T_{10}^I$, meaning the contribution of $O_p$ versus $D_p$ in each set of directions, is a priori not settled (or at least that of its integral, as we implicitly mean here). The sum of all of them, $T_{10}$, has to be positive though (see below \eqref{R4T10F}). This differs with respect to the case of a BPS-like configuration. Let us assume for now that $\forall\, I,\ T_{10}^I \geq 0,\ {\rm and} \ \exists\, I \ {\rm s.t.}\ T_{10}^I > 0$. By definition, for $J\neq I$, one has in addition $0 \leq \delta_{a_{||_I}}^{a_{||_J}} < \delta_{a_{||_I}}^{a_{||_I}} = p-3$. In that case, we deduce
\beq
0 \leq  \sum_I \sum_{J\neq I} \delta_{a_{||_I}}^{a_{||_J}} T_{10}^J  < (p-3) (N-1) T_{10} \ .\label{inequality}
\eeq
This shows that the source terms are then appropriate contributions to the inequalities \eqref{require3}.} So all these differences make it simpler to satisfy the inequalities \eqref{require3}: this may explain why only de Sitter solutions with intersecting sources, i.e.~$N>1$, are known.

The requirement \eqref{require3} is conceptually interesting but remains cumbersome for a practical use, due to the term related to the overlap of sources. Despite various attempts with this term, we did not reach much refined constraints, except in a particular case of overlap that we now focus on.

\subsection{The case of homogeneous overlap}\label{sec:homoover}

There is an interesting situation where sources in each set $I$ overlap with all others in the same manner. We call this symmetric situation an ``homogeneous overlap'', and define it as follows:
\bea
&\mbox{{\bf Homogeneous overlap assumption}:} \label{assumptionoverlap}\\
& \quad \mbox{Each set $I$ overlaps all others in the same amount,} \nn\\
& \quad \mbox{meaning}\ \forall I,\, J\neq I,\ \delta_{a_{||_I}}^{a_{||_J}} = N_o\ \mbox{independent of}\ I,\, J\ .\nn
\eea
By definition, this number of overlapping directions $N_o$ is such that $0 \leq N_o < p-3$; it includes the case of no overlap. Strictly speaking, $N_o$ is not necessarily an integer, e.g.~in the case of sources at angles smaller than $\tfrac{\pi}{2}$. But one should in general be able to introduce projections towards an orthogonal basis and thus avoid this subtlety. All known de Sitter solutions with fixed $p$, namely $p=6$ in \cite{Danielsson:2011au}, as well as all known Minkowski solutions on solvmanifolds with intersecting sources ($p=5,6$, see Section \ref{sec:susymink}), verify the assumption \eqref{assumptionoverlap}. All these solutions admit in addition the particular value $N_o=p-5$:\footnote{\label{foot:dSorbif}The internal geometry of de Sitter solutions of \cite{Danielsson:2011au} is built with one $O_6$ involution combined with orbifold actions, acting on a group manifold. Let us detail how one gets to the picture of $N=4$ intersecting sets of $O_6$ with $N_o=1$. Four cases are considered. The first one called ``standard $\mathbb{Z}_2 \times \mathbb{Z}_2$ orientifold'' is analogous to the known $T^6/\mathbb{Z}_2 \times \mathbb{Z}_2$: the $O_6$ is along $e^1\w e^2 \w e^3$, and the orbifold actions combined to the orientifold involution are equivalent to no orbifold but three other $O_6$, along $e^3\w e^4 \w e^6$, $e^2\w e^5 \w e^6$, and $e^1\w e^4 \w e^5$. For the ``non-standard $\mathbb{Z}_2 \times \mathbb{Z}_2$ orientifold'', one proceeds similarly and looks for invariant three-forms under the combinations of the involution and orbifold actions, which correspond to the internal spaces wrapped by the $O_6$: several choices are possible, as different base choices for these various $O_6$, one being $e^1\w (e^2+e^3) \w (e^5-e^6)$, $e^4\w (e^2+e^3) \w (e^2-e^3)$, $e^4\w (e^5+e^6) \w (e^5-e^6)$, $e^1\w (e^2-e^3) \w (e^5+e^6)$. The last two cases consider $\mathbb{Z}_3$ extensions into a non-abelian orbifold of the previous cases, where the $\mathbb{Z}_2 \times \mathbb{Z}_2$ remains a subgroup of the orbifold group. So the configuration of $O_6$ remains, it simply gets orbifolded. In particular for the ``standard orientifold'' case, the three-form $e^1\w e^2 \w e^3$ remains invariant under the new $\mathbb{Z}_3$ action, etc. We conclude that all de Sitter solutions of \cite{Danielsson:2011au} have $N=4$ intersecting sets of $O_6$ with $N_o=1$.} we will see that this special value comes out naturally from the equations. The assumption of homogeneous overlap is thus motivated by a technical simplification but also by the known examples. Intuitively, such a symmetric configuration of sources would also be thought to preserve some off-shell supersymmetry and bring stability; but this is difficult to verify in full generality.

We now focus on this case and make use of \eqref{assumptionoverlap}: it allows to factorize $N_o$ and build $T_{10}$
\beq
\sum_I \sum_{J\neq I} \delta_{a_{||_I}}^{a_{||_J}} T_{10}^J = N_o \sum_I \sum_{J\neq I} T_{10}^J = N_o (N-1) T_{10} \ . \label{equalityT10}
\eeq
Inserting this in \eqref{finalbis}, and replacing $T_{10}$ through \eqref{R4T10F}, we obtain
\bea
\boxed{\mbox{Result:}}\quad & ((N-1)(p-3 -N_o) + 2) {\cal R}_4 \label{finalhomo}\\
& = 2 \varepsilon_p e^{\phi} \sum_I (\d F_k^{(0)_I})_{\bot_I} - \sum_I  2 e^{2\phi} |F_k^{(0)_I}|^2  - \sum_I \left|*_{\bot_I}H|_{\bot_I} + \varepsilon_p e^{\phi} F_{k-2}|_{\bot_I} \right|^2 \nn \\
& - \sum_I \sum_{a_{||_I}} \left| *_{\bot_I}( \d e^{a_{||_I}})|_{\bot_I} - \varepsilon_p e^{\phi}\, \iota_{a_{||_I}} F_k^{(1)_I} \right|^2 \nn\\
& + \sum_I (- 2{\cal R}_{||_I} - 2 {\cal R}_{||_I}^{\bot_I}  + |H^{(2)_I}|^2 + 2 |H^{(3)_I}|^2 )  \nn\\
& - (N-1) (p-5-N_o) e^{2\phi}( |F_{k-4}|^2 + |F_{k-2}|^2 + |F_{k}|^2 ) \nn\\
& \hspace{-1.3in} - e^{2\phi} \bigg( ((N-1)(p-3 - N_o) +2) |F_{k+2}|^2 + ((N-1)(2p-10-N_o)+ p-5) |F_{k+4}|^2 + (3-N_o)(N-1) |F_{k+6}|^2 \bigg) \nn\\
& \hspace{-1.3in} + e^{2\phi} \sum_I \bigg( \frac{1}{2}  \big( |(*_6 F_{5})|_{\bot_I}|^2 - |F_{5}|_{\bot_I}|^2 \big) + \sum_{n=2}^{p-3} (n-1) \big( |F_{k+2}^{(n)_I}|^2 + \frac{p-6}{2} |F_{k+4}^{(n)_I}|^2 + \frac{p-7}{4} |F_5^{(n)_I}|^2  \big) \bigg) \nn  \ .
\eea
The coefficient on the left-hand side is again strictly positive. Let us detail the last two lines with fluxes:
\bea
& p=3:\ - e^{2\phi} ( \mbox{fluxes} ) = 0 \\
& p=4:\ - e^{2\phi} ( \mbox{fluxes} ) = - ((N-1)(1- N_o) +2) e^{2\phi} |F_{6}|^2 \nn\\
& p=5:\ - e^{2\phi} ( \mbox{fluxes} ) = - e^{2\phi} ((N-1)(2- N_o) +2) |F_5|^2  + \frac{1}{2} e^{2\phi} \sum_I (|(*_6 F_{5})|_{\bot_I}|^2 + |F_5^{(2)_I}|^2 )   \nn\\
& p=6:\ - e^{2\phi} ( \mbox{fluxes} ) = - e^{2\phi} ((N-1)(3- N_o) +2) |F_{4}|^2 + e^{2\phi} \sum_I  (  |F_4^{(2)_I}|^2 + 2 |F_4^{(3)_I}|^2) \nn\\
& \phantom{p=6:\ - e^{2\phi} ( \mbox{fluxes} ) =} - e^{2\phi} ((N-1)(2- N_o) +1) |F_{6}|^2 \nn\\
& p=7:\ - e^{2\phi} ( \mbox{fluxes} ) = - e^{2\phi} ((N-1)(4- N_o) +2) |F_{3}|^2 + e^{2\phi} \sum_I  (  |F_3^{(2)_I}|^2 + 2 |F_3^{(3)_I}|^2) \nn\\
& \phantom{p=7:\ - e^{2\phi}} - e^{2\phi} ((N-1)(4- N_o) +2) |F_5|^2  + \frac{1}{2} e^{2\phi} \sum_I \big( |(*_6 F_{5})|_{\bot_I}|^2 + \sum_{n= 2}^{4} (n-1) |F_5^{(n)_I}|^2 \big) \nn\\
& p=8:\ - e^{2\phi} ( \mbox{fluxes} ) = - e^{2\phi} \Big( ((N-1)(5- N_o) +2) |F_{2}|^2 - \sum_I |F_{2}^{(2)_I}|^2 \nn\\
& \phantom{p=8:\ - e^{2\phi}} + ((N-1)(6- N_o) +3) |F_{4}|^2 - \sum_I \sum_{n=2}^4 (n-1) |F_{4}^{(n)_I}|^2  + (3 -N_o)(N-1) |F_{6}|^2 \Big) \ .\nn
\eea
These contributions are all $\leq 0$, provided $p=4$, or $p\geq 5$ and $0\leq N_o \leq p-5$.\footnote{For the case $p=5$ with $0\leq N_o \leq p-4$, i.e.~bounding $N_o$ with its maximal integer value, these fluxes contributions are $\leq 0$. The cases $p=6,7,8$ however require the further restriction $0\leq N_o \leq p-5$.} In addition, the other flux term in \eqref{finalhomo} proportional to $(N-1)$ points towards the same bound. We deduce the following requirement:
\bea
\boxed{\mbox{Result:}} & \quad \mbox{For $p\geq 5$ with $0\leq N_o \leq p-5$, having a de Sitter solution requires} \label{requireNo}\\
& \ \int_6 {\rm vol}_6 \bigg( - \sum_I \sum_{a_{||_I}} |(\d e^{a_{||_I}})|_{\bot_I}|^2 - e^{\phi} (N-1) \frac{(p-3-N_o)}{p+1} T_{10} \bigg) \nn\\
& \qquad \qquad \qquad < \int_6 {\rm vol}_6  \sum_I ( 2{\cal R}_{||_I} + 2 {\cal R}_{||_I}^{\bot_I}  - |H^{(2)_I}|^2 - 2 |H^{(3)_I}|^2 )  < 0 \ , \nn
\eea
which can be turned into a no-go theorem. The left inequality is obtained from \eqref{require3}. The resulting constraint on the combination of curvature terms and $H$-flux components is more interesting than before.\\

We have identified a set of parameters, namely $p\geq 5$ and $0\leq N_o \leq p-5$, for which we reached interesting constraints on de Sitter solutions. Let us discuss in more details this range of parameters. For $p=8$ in a six-dimensional compact manifold, having at least $N=2$ sets requires $N_o \geq 4$, due to the dimensionality of the objects. This does not fit in $N_o \leq p-5$, so $p=8$ can only have $N_o=0$, implying $N=1$. A similar reasoning indicates for $p=7$ that $N_o=2$ at least, i.e.~$N_o = p-5$; otherwise $N_o=0$ and $N=1$. For $p=5$, one is forced to take $N_o=0$, allowing then $N=1,2,3$. Finally for $p=6$, one has $0\leq N_o \leq 1$: the case $N_o=0$ imposes $N=1,2$ while $N_o=1$ gives more possibilities for $N$. To summarize, restricting ourselves to intersecting sources and an integer $N_o$, the range of parameters for which we obtain the interesting constraints \eqref{requireNo} is
\bea
& p\geq 5,\ 0\leq N_o \leq p-5,\ N\geq2,\ N_o\ \mbox{is an integer} \label{parametersNo}\\
& \Rightarrow (p=5, N_o=0, N=2,3),\ (p=6, N_o=0, N=2),\ (p=6, N_o=1, N),\ (p=7, N_o=2, N) \ .\nn
\eea
This set of values will be of particular interest when discussing Minkowski solutions in Section \ref{sec:susymink}. For de Sitter solutions, the requirement \eqref{requireNo} only improves the one without homogeneous overlap, \eqref{require3}, in the case $p=6$ and $N_o=1$ (for $N_o=0$, the two are identical). Interestingly though, this is precisely the case of the known de Sitter solutions \cite{Danielsson:2011au} (see footnote \ref{foot:dSorbif} for details). The requirement \eqref{requireNo} should be especially useful when completing the present work with a study of the solutions (meta)stability.

\subsection{On group manifolds}\label{sec:groupmanif}

We focus here on the particular case where $\mmm$ is a compact group manifold (see e.g.~a list in \cite{Danielsson:2011au}). All known classical de Sitter solutions were obtained on such manifolds. Interestingly, one then obtains more constraints on the curvature terms ${\cal R}_{||_I}$ in \eqref{Rpar} and ${\cal R}_{||_I}^{\bot_I}$ in \eqref{Rbotpar}, that play a crucial role in our conditions \eqref{requirep=4}, \eqref{require2}, \eqref{require3} and \eqref{requireNo} for de Sitter solutions. Indeed, on such manifolds, the $f^a{}_{bc}$ are constant, making any orientifold projection more constraining. Assuming in the following that there is an orientifold in each set $I$ (see footnote \ref{foot:signT10I} on this point), the compatibility of its projection with the algebra or geometry requires
\beq
f^{a_{\bot_I}}{}_{b_{\bot_I}c_{\bot_I}}= f^{a_{\bot_I}}{}_{b_{||_I}c_{||_I}}= f^{a_{||_I}}{}_{b_{||_I}c_{\bot_I}}=0 \ .
\eeq
This makes ${\cal R}_{||_I}^{\bot_I}$ reduce to the first line of \eqref{Rbotpar}, that we rewrite as follows
\beq
2 {\cal R}_{||_I}^{\bot_I} =  - \frac{1}{2} |\delta_{d_{\bot_I}a_{\bot_I}} f^{a_{\bot_I}}{}_{b_{\bot_I}c_{||_I}} + \delta_{b_{\bot_I}a_{\bot_I}} f^{a_{\bot_I}}{}_{d_{\bot_I}c_{||_I}} |^2=  - 2 |\delta_{a_{\bot_I}(d_{\bot_I}} f^{a_{\bot_I}}{}_{b_{\bot_I})c_{||_I}} |^2 \leq 0 \ ,\label{Rparbotsquare}
\eeq
where the square is obtained by the contraction (without any factor here) of the three indices of this tensor with the flat metric, hence the sign. We also rewrite ${\cal R}_{||_I}$ in a similar fashion, although less constraining
\bea
2 {\cal R}_{||_I}  =& \ -\delta^{ab} f^{d_{||_I}}{}_{c_{||_I}a_{||_I}} f^{c_{||_I}}{}_{d_{||_I}b_{||_I}}- \frac{1}{2} \delta^{ch}\delta^{dj}\delta_{ab} f^{a_{||_I}}{}_{c_{||_I}j_{||_I}} f^{b_{||_I}}{}_{h_{||_I}d_{||_I}} \\
=& \ -\frac{1}{2}\delta^{ab} f^{d_{||_I}}{}_{c_{||_I}a_{||_I}} f^{c_{||_I}}{}_{d_{||_I}b_{||_I}}- |\delta_{a_{||_I}(d_{||_I}} f^{a_{||_I}}{}_{b_{||_I})c_{||_I}} |^2 \nn\\
=& \ - 2 |\delta_{a_{||_I}(d_{||_I}} f^{a_{||_I}}{}_{b_{||_I})c_{||_I}} |^2 + \frac{1}{2} |f^{a_{||_I}}{}_{c_{||_I}j_{||_I}} |^2 \nn
\eea
where again, exceptionally, we do not include any factor in the squares and simply contract all free indices. On some solvmanifolds, the two terms in these expressions can cancel each other, leading to a vanishing ${\cal R}_{||_I}$, while on others, it tends to be negative. On nilmanifolds, the product $f^{d_{||_I}}{}_{c_{||_I}a_{||_I}} f^{c_{||_I}}{}_{d_{||_I}b_{||_I}} =0$ necessarily, giving
\beq
{\cal R}_{||_I} \leq 0 \quad \mbox{on nilmanifolds} \ .
\eeq
These expressions and signs of the curvature terms are very interesting in view of the condition \eqref{requireNo}: for instance, on nilmanifolds, the sum of curvature terms is automatically negative as long as only one of ${\cal R}_{||_I}$ or ${\cal R}_{||_I}^{\bot_I}$ is non-zero. This also gives an idea on typical signs, and gives ways to compute these curvature terms. Let us add a word on the $H$-flux: it should in general be odd under the orientifold involution. If one restricts to a constant $H$-flux, as often done when looking for solutions on group manifolds, then components $H^{(1)}, H^{(3)}$ have to vanish, leaving only $H^{(2)}$ in \eqref{requireNo}.

If in addition sources do not overlap (e.g.~the previous $N_o=0$), one necessarily has for $I\neq J$
\beq
f^{a_{||_I}}{}_{b_{||_I}c_{||_I}} = f^{a_{\bot_J}}{}_{b_{\bot_J}c_{\bot_J}} = 0 \Rightarrow {\cal R}_{||_I} = 0 \ .\label{Rpar=0}
\eeq
Furthermore, for ($N=2$) non-overlapping $O_6$, one gets
\beq
f^{a_{||_I}}{}_{b_{\bot_I}c_{\bot_I}} = f^{a_{\bot_J}}{}_{b_{||_J}c_{||_J}} = 0 \ ,\ f^{a_{\bot_I}}{}_{b_{\bot_I}c_{||_I}} = f^{a_{||_J}}{}_{b_{||_J}c_{\bot_J}} = 0 \ .
\eeq
We infer that non-overlapping $O_6$ on a group manifold forces all $f^a{}_{bc}=0$, i.e.~the manifold is restricted to be a torus. Similarly, one verifies that $N=2$ non-overlapping $O_6$ impose $H^{(2)_I}= H^{(1)_J}$, which vanishes for a constant flux. From \eqref{requireNo}, we deduce
\bea
\boxed{\mbox{Result:}} & \quad \mbox{There is no de Sitter solution with non-overlapping $O_6$} \label{nogo6group} \\
& \quad \mbox{on a group manifold (with constant $H$-flux).} \nn
\eea
This is interesting, given this was one case of \eqref{parametersNo}. For non-overlapping $O_5$, we cannot reach such a constraint. The $f^a{}_{bc}$ can be related from one set $I$ to another set $J$, but it does not set all of them to zero. Rather, the condition \eqref{Rpar=0} indicates that the sum of curvature terms is automatically negative, as preferred in \eqref{requireNo}, as long as only one ${\cal R}_{||_I}^{\bot_I}$ is non-zero.\\

We finally comment on $O_4$. On group manifolds with orientifold, one is left with three types of structure constants, $f^{a_{||}}{}_{b_{||}c_{||}}$, $f^{a_{||}}{}_{b_{\bot}c_{\bot}}$ or $f^{a_{\bot}}{}_{b_{\bot}c_{||}}$; for $p=4$, the first one vanishes since there is only one parallel direction. This means that each non-zero $f^a{}_{bc}$ carries exactly one index parallel to one $O_4$. Because $f^a{}_{bc}$ have three indices, there can be at most $N=3$ sets with $O_4$. If there are more (necessarily non-overlapping) $O_4$, i.e.~$N\geq 4$, all structure constants must vanish, and the manifold is then a torus. Indeed, the non-zero $f^a{}_{bc}$ have indices at most along three $O_4$, and those appear for a fourth one as $f^{a_{\bot}}{}_{b_{\bot}c_{\bot}}$ which then vanishes. This explains why there are very few nilmanifolds and solvmanifolds that admit $O_4$ in \cite{Grana:2006kf}. However, even on a torus, it is difficult (see Section \ref{sec:p=4}) to reach any conclusion regarding de Sitter or Minkowski solutions with intersecting $p=4$ sources.

\section{Towards Minkowski solutions}\label{sec:susymink}

In this section, the expressions derived for ${\cal R}_4$ are used to look for Minkowski solutions with intersecting sources, and new expressions are developed.

\subsection{Foreword}\label{sec:foreword}

In the case of parallel sources \cite{Andriot:2016xvq}, i.e.~$N=1$, the formulas derived, in particular the ${\cal R}_4$ expression analogue to \eqref{finalbis} or \eqref{finalhomo}, have been used in \cite{Andriot:2016ufg} to find a class of Minkowski solutions. Those are not necessarily supersymmetric, but include supersymmetric solutions, and are inspired by them. To obtain this class, the key point in \cite{Andriot:2016ufg} has been to set to zero the curvature terms $ {\cal R}_{||} +  {\cal R}_{||}^{\bot} $ as well as the components $H^{(2)}, H^{(3)}$, through an assumption. All other terms in the ${\cal R}_4$ expression had the same definite sign, so asking for ${\cal R}_4=0$ would then set each term separately to zero, providing this way an interesting ansatz of solution.

We aim here at a similar result with intersecting sources, i.e.~$N\geq 2$. We focus in the following on the case of homogeneous overlap \eqref{assumptionoverlap} with $p\geq 5$ and $0\leq N_o \leq p-5$, which restricts to $p=5,6,7$. For an integer $N_o$, this eventually corresponds to the finite set of parameters \eqref{parametersNo}, but we do not assume this for now. As explained in Section \ref{sec:homoover}, considering $0\leq N_o \leq p-5$ gives all terms in the ${\cal R}_4$ expression \eqref{finalhomo} the same definite sign (upon integration), except for the curvature terms and specific $H$-flux components: this is analogous to the case of parallel sources. We then assume that these terms vanish, i.e.~$\sum_I {\cal R}_{||_I} +  {\cal R}_{||_I}^{\bot_I} = 0$ and $\forall I,\ |H^{(2)_I}|^2 = |H^{(3)_I}|^2=0$. As a consequence, each of the other terms, integrated, should vanish with ${\cal R}_4=0$, and so should each integrand. In particular, $\forall I,\ F_k^{(0)_I} = 0$ (this is due to the absence of warp factor), which sets to zero the total derivatives in \eqref{finalhomo}. All terms with definite sign in \eqref{finalhomo} then vanish. To avoid a trivial solution,\footnote{If $N_o \neq p-5$, it restricts the parameters to only $p=6$ with $0 \leq N_o < 1$. We deduce from \eqref{finalhomo} that all RR fluxes vanish. From \eqref{4dtracefinalFk}, we deduce as well that $H=0$, from \eqref{R4T10F} that $T_{10}$=0, and from \eqref{4dtracesansT10} that ${\cal R}_6=0$. From the four-dimensional Einstein equation, we also get ${\cal R}_{\mu\nu}= T_{\mu\nu}=0$, leaving us with quite trivial solutions. Some non-triviality might be recovered with warp factors, similarly to the solutions with $N=1$ described in \cite{Andriot:2015sia}.} one is forced to set $N_o = p-5$. The other terms lead to the following ansatz:
\bea
& \forall I,\ F_k^{(0)_I}= 0,\ \iota_{a_{||_I}} F_k^{(1)_I}= \varepsilon_p e^{-\phi} *_{\bot_I}( \d e^{a_{||_I}})|_{\bot_I} \ ,\label{ansatz0}\\
& F_{k-2}|_{\bot_I} = - \varepsilon_p e^{-\phi} *_{\bot_I}H|_{\bot_I} \ , \nn
\eea
while the remaining RR fluxes require more focus. For these values of $p$, $F_{k-4}=F_{k+6}=0$, and $F_k= F_k^{(1)_I} + F_k^{(2)_I}\ \forall I$. For $p=5$, \eqref{finalhomo} sets $F_{k+2}=F_{k+4}=0$, but some components can remain for $p=6,7$. These two fluxes are however also set to zero combining the above ansatz with \eqref{4dtracefinalFk}. So overall, we get $F_{k-4}=F_{k+2}=F_{k+4}= F_{k+6} = 0$. With the above ansatz and \eqref{dFrewrite}, one shows in addition $\forall I,\ -  2 \varepsilon_p e^{\phi} (\d F_k)_{\bot_I} + 2 e^{2\phi} |F_k^{(1)_I}|^2 = 0$. We deduce with \eqref{firstsquaregen}
\beq
2 e^{2\phi} ( |F_{k}|^2 -  \sum_I |F_k^{(1)_I}|^2 )  +|H|^2 - \sum_I |H|_{\bot_I}|^2 + e^{2\phi} (  |F_{k-2}|^2 - \sum_I |F_{k-2}|_{\bot_I}|^2 ) = 0 \ .
\eeq
Except with a fluxless solution, such a cancelation looks very unlikely. There is thus a problem with this solution ansatz.\\

To understand better the situation, let us look at explicit examples. From the list in \cite{Andriot:2015sia} (see also the one in \cite{Andriot:2010sya}), we read all known Minkowski supersymmetric solutions on solvmanifolds with intersecting sources, not considering the torus. Those are:
\begin{itemize}
  \item two solutions with $N=2$ $O_5$, $N_o=0$, on the nilmanifold $n 3.14$ (same directions of the sources for both solutions) in \cite{Grana:2006kf, Andriot:2008va};
  \item several solutions with $N=2$ $O_5$, $N_o=0$, and with $N=2$ $O_6$, $N_o=1$, on the solvmanifold $s 2.5$ (for each $p$, there are two different possible sets of directions for the sources on the manifold, related to each other by a symmetry of the algebra) in \cite{Camara:2005dc, Grana:2006kf, Andriot:2008va};
  \item one solution with $N=2$ $O_6$, $N_o=1$, on the solvmanifold of algebra $\mathfrak{g}_{5.7}^{1,-1,-1} \oplus \mathbb{R}$, the hyperbolic counterpart of $s 2.5$ (two different possible sets of directions of the sources) in \cite{Camara:2005dc};
  \item one solution with $N=2$ $O_6$, $N_o=1$, on the solvmanifold of algebra $\mathfrak{g}_{5.17}^{q,-q,r} \oplus \mathbb{R} \approx s 2.5 + q (\mathfrak{g}_{5.7}^{1,-1,-1} \oplus \mathbb{R}) $, i.e.~a combination of the previous two, in \cite{Andriot:2010ju}.
\end{itemize}
First, we note that all these examples are obtained for $N=2$, $p=5,6$ and $N_o=p-5$. We now look at the curvature terms: because these are group manifolds, we can use the results of Section \ref{sec:groupmanif}. Considering the examples of \cite{Grana:2006kf}, one verifies that $f^{a_{||_I}}{}_{b_{||_I}c_{||_I}} =0$, giving ${\cal R}_{||_I}=0$. In addition, one obtains for these solutions thanks to \eqref{Rparbotsquare} that $\exists I$ such that ${\cal R}_{||_I}^{\bot_I} \neq 0$. We conclude with \eqref{Rparbotsquare} that $\sum_I {\cal R}_{||_I} +  {\cal R}_{||_I}^{\bot_I} < 0$. More generally, we verify that all above solutions have non-zero curvature terms. In addition, the first solution of \cite{Andriot:2008va} admits non-zero $H^{(2)_I}$ (as well as $H^{(0)_I}$). We conclude that our initial assumptions, analogous to the case of parallel sources, do not work!

The various curvature terms ${\cal R}_{||_I},\,  {\cal R}_{||_I}^{\bot_I}$, are non-zero if some corresponding submanifolds of $\mmm$ are curved. The bigger $N$ is, the more submanifolds are probed, therefore the more probable it is to have one non-vanishing curvature term. As explained for group manifolds, one non-zero curvature term can be enough to have $\sum_I {\cal R}_{||_I} +  {\cal R}_{||_I}^{\bot_I} $ non-vanishing. So this gives an intuitive explanation why one should not expect vanishing curvature terms with intersecting sources. If we now come back to \eqref{finalhomo}, one may wonder what compensates these curvature terms (and the $H$-flux components) for a Minkowski solution. In the examples of \cite{Grana:2006kf}, the only non-zero flux is $F_k$, and one verifies that $F_k=F_k^{(1)_I}\ \forall I$. The only way to compensate the curvature terms is thus a violation of the ansatz
\beq
\iota_{a_{||_I}} F_k^{(1)_I}= \varepsilon_p e^{-\phi} *_{\bot_I}( \d e^{a_{||_I}})|_{\bot_I}\ \Leftrightarrow \  F_k^{(1)_I}= \varepsilon_p e^{-\phi} \delta_{a_{||_I} b_{||_I}} e^{a_{||_I}} \w *_{\bot_I}( \d e^{b_{||_I}})|_{\bot_I}  \ .\label{pseudoBPSlike}
\eeq
This is indeed what happens in these examples!\\

The ansatz \eqref{pseudoBPSlike} is a valid one for $N=1$. As argued in \cite{Andriot:2016xvq, Andriot:2016ufg}, it should be understood as coming from the calibration condition of the sources. In the supersymmetric case, this condition even boils down to one of the supersymmetry conditions \cite{Koerber:2005qi, Martucci:2005ht, Koerber:2007jb}, but we remain more general here. We then follow this idea of the calibration condition to deduce an appropriate ansatz with intersecting sources. We mimic the derivation of this condition as done in Appendix B of \cite{Andriot:2016xvq}, in the case of multiple sources, and obtain, without warp factor and with constant dilaton,
\beq
F_{k} = (-1)^{p} \varepsilon_p e^{-\phi} *_6 \d\big( \sum_J {\rm vol}_{||_J} \big) \ ,\quad 0 \leq k=8-p \leq 5 \ .\label{calibflux}
\eeq
Using $f^{a_{||_J}}{}_{a_{||_J}b_{\bot_J}}=0$, equivalent to \eqref{trace} given the compactness of $\mmm$, the above can be rewritten as
\bea
F_{k} & = (-1)^{p} \varepsilon_p e^{-\phi} *_6 \big( \sum_J \sum_{a_{||_J}} ( \d e^{a_{||_J}})|_{\bot_J} \w  \iota_{a_{||_J}} {\rm vol}_{||_J} \big) \nn \\
& = \varepsilon_p e^{-\phi} \sum_J \delta_{a_{||_J} b_{||_J}} e^{a_{||_J}} \w *_{\bot_J} ( \d e^{b_{||_J}})|_{\bot_J}  \ ,\label{BPSlikemultiple}
\eea
where the sum on $a_{||_J}$ is traded for $\delta_{a_{||_J} b_{||_J}}$. The difference with \eqref{pseudoBPSlike} is clear: $F_k^{(1)_I}$ gets contributions not only from the $I$ term in the sum, corresponding to \eqref{pseudoBPSlike}, but also from other terms $J\neq I$ if relations like $a_{||_I}=a_{\bot_J}$, $a_{\bot_I}=a_{||_J}$ hold. This allows $F_k=F_k^{(1)_I}\ \forall I$.\\

To summarize, the ${\cal R}_4$ expression \eqref{finalhomo} does not provide an appropriate ansatz for Minkowski solutions, on the contrary to the case of parallel sources \cite{Andriot:2016xvq, Andriot:2016ufg}. One reason is that non-trivial Minkowski solutions with intersecting sources rather admit non-vanishing curvature terms. This led us to propose another ansatz for the sourced flux $F_k$, inspired by the idea of calibration condition. We now derive a new expression for ${\cal R}_4$ where this ansatz appears.

\subsection{New derivation and comments}\label{sec:derivcomment}

To derive an expression for ${\cal R}_4$ where the $F_k$ ansatz \eqref{calibflux} or \eqref{BPSlikemultiple} enters, one should bring the sum on $J$ in \eqref{BPSlikemultiple} inside the square of the BPS-like condition, instead of outside as in \eqref{finalhomo}. We do so in the following by revisiting the treatment of the Bianchi identity. One could do the same for the $H$-flux and $F_{k-2}$ term: we tackle this in Appendix \ref{ap:HF}. The new treatment of the Bianchi identity goes as follows:
\bea
\sum_I (\d F_k)_{\bot_I} &= \sum_I *_{\bot_I} (\d F_k)|_{\bot_I} = \sum_I *_6 ({\rm vol}_{||_I}\w (\d F_k)|_{\bot_I}  )\\
& = \sum_I *_6 ({\rm vol}_{||_I}\w \d F_k) = \sum_I *_6 ({\rm vol}_{||_I}\w \d F_k^{(0)_I}) + \sum_I *_6 ({\rm vol}_{||_I}\w \d \sum_{(n)_I>0} F_k^{(n)_I}) \nn\\
& =  (-1)^{p-1} \sum_I *_6 \d ( {\rm vol}_{||_I}\w  F_k^{(0)_I}) + (-1)^p \sum_I *_6 (\d{\rm vol}_{||_I}\w F_k^{(0)_I}) \nn\\
& \phantom{(-1)^{p-1} \sum_I *_6 \d ( {\rm vol}_{||_I}\w  F_k^{(0)_I})} + (-1)^p \sum_I *_6 ( \d{\rm vol}_{||_I}\w \sum_{(n)_I>0} F_k^{(n)_I}) \nn\\
& =  (-1)^{p-1} \sum_I *_6 \d ( {\rm vol}_{||_I}\w  F_k^{(0)_I}) + (-1)^p \sum_I *_6 (\d{\rm vol}_{||_I}\w F_k) \ , \nn
\eea
from which we get
\bea
& -2\varepsilon_p e^{\phi} \sum_I (\d F_k)_{\bot_I} + (-1)^{p-1} 2\varepsilon_p e^{\phi} \sum_I *_6 \d ( {\rm vol}_{||_I}\w  F_k^{(0)_I}) \label{dFnew}\\
& = -(-1)^p 2\varepsilon_p  *_6 \Big(\d \big( \sum_I {\rm vol}_{||_I} \big) \w e^{\phi} F_k \Big) \nn\\
& = -(-1)^p \varepsilon_p *_6 \Big(e^{\phi} F_k \w *_6^2 \d \big( \sum_I {\rm vol}_{||_I} \big) \Big) - (-1)^p \varepsilon_p e^{\phi} *_6 \Big( *_6 \d \big( \sum_I {\rm vol}_{||_I} \big) \w *_6 e^{\phi} F_k \Big) \nn\\
& = \left|(-1)^{p} \varepsilon_p *_6 \d\big( \sum_I {\rm vol}_{||_I} \big) - e^{\phi} F_k \right|^2 - |\d\big( \sum_I {\rm vol}_{||_I} \big) |^2 - e^{2\phi} |F_k|^2 \ , \nn
\eea
and we refer to \eqref{blasquare} or \cite{Andriot:2016xvq} for more details on the signs. We recall from \eqref{calibflux} and \eqref{BPSlikemultiple} that
\beq
(-1)^{p}  *_6 \d\big( \sum_I {\rm vol}_{||_I} \big) = \sum_I \delta_{a_{||_I} b_{||_I}} e^{a_{||_I}} \w *_{\bot_I} ( \d e^{b_{||_I}})|_{\bot_I}  \ .
\eeq
We deduce
\bea
& |\d\big( \sum_I {\rm vol}_{||_I} \big) |^2 = | \sum_I \delta_{a_{||_I} b_{||_I}} e^{a_{||_I}} \w *_{\bot_I} ( \d e^{b_{||_I}})|_{\bot_I} |^2 = \sum_I \sum_{a_{||_I}} | ( \d e^{a_{||_I}})|_{\bot_I} |^2 + \sum_{I\neq J} {\cal O}_{IJ} \ , \label{squaredvol}\\
& {\cal O}_{IJ} = *_6\left(  \delta_{a_{||_I} b_{||_I}} e^{a_{||_I}} \w *_{\bot_I} ( \d e^{b_{||_I}})|_{\bot_I} \w *_6 \big(  \delta_{a_{||_J} b_{||_J}} e^{a_{||_J}} \w *_{\bot_J} ( \d e^{b_{||_J}})|_{\bot_J} \big) \right) \ ,\nn
\eea
where ${\cal O}_{IJ} = {\cal O}_{JI}$ so $\sum_{I\neq J} {\cal O}_{IJ} = 2 \sum_{I < J} {\cal O}_{IJ}$. We have brought the sum on $I$ inside the square, allowing to make the ansatz \eqref{calibflux} appear in \eqref{dFnew} through the square of a BPS-like condition. The cost of having a sum inside a square is that it leads to double product terms, given here by ${\cal O}_{IJ}$ which are not easy to evaluate in full generality. Still, starting from \eqref{firstsquaregen}, we deduce from these results
\bea
{\cal R}_4  = -\frac{2}{p+1} \bigg( & (-1)^{p} 2\varepsilon_p e^{\phi} \sum_I *_6 \d ( {\rm vol}_{||_I}\w  F_k^{(0)_I}) - |\d\big( \sum_I {\rm vol}_{||_I} \big) |^2 \label{newR41}\\
& + \left|(-1)^{p} \varepsilon_p *_6 \d\big( \sum_I {\rm vol}_{||_I} \big) - e^{\phi} F_k \right|^2 + \sum_I \left|*_{\bot_I}H|_{\bot_I} + \varepsilon_p e^{\phi} F_{k-2}|_{\bot_I} \right|^2  \nn \\
& + \sum_I (|H|^2- |H|_{\bot_I}|^2) + e^{2\phi}  \sum_I ( |F_{k-2}|^2 - |F_{k-2}|_{\bot_I}|^2 ) -(N-1) (|H|^2 + e^{2\phi} |F_{k-2}|^2 )  \nn\\
& + e^{2\phi} ( |F_{k}|^2 + 3 |F_{k+2}|^2 + 4 |F_{k+4}|^2 + 5 |F_{k+6}|^2 )  \bigg)  \ . \nn
\eea

We then proceed as in Section \ref{sec:derivgen}: with the internal trace \eqref{tracepargen}, we obtain
\bea
& - |\d\big( \sum_I {\rm vol}_{||_I} \big) |^2  + \sum_I (|H|^2- |H|_{\bot_I}|^2) + e^{2\phi}  \sum_I ( |F_{k-2}|^2 - |F_{k-2}|_{\bot_I}|^2 ) \nn\\
= &  - \sum_{I\neq J} {\cal O}_{IJ} + 2 \sum_I ({\cal R}_{||_I} +   {\cal R}_{||_I}^{\bot_I}) \nn\\
& +\sum_I\Bigg( -  \frac{p-3}{2}  {\cal R}_4 - e^{\phi} \left( T_{a_{||_I}}^{a_{||_I}}  - \frac{p-3}{p+1} T_{10} \right) \nn\\
& \phantom{+\sum_I\Bigg(} - e^{2\phi} \bigg(  |F_{k}|^2 - |F_{k}|_{\bot_I}|^2 + |F_{k+2}|^2 + (9-p) |F_{k+4}|^2 + 5 |F_{k+6}|^2  + \frac{1}{2} ( |(*_6 F_{5})|_{\bot_I}|^2 - |F_{5}|_{\bot_I}|^2 ) \bigg) \nn\\
& \phantom{+\sum_I\Bigg(} - \sum_{n=2}^{p-3} (n-1) \left(|H^{(n)_I}|^2 + e^{2\phi} (  |F_k^{(n)_I}|^2 + |F_{k+2}^{(n)_I}|^2 + \frac{p-6}{2} |F_{k+4}^{(n)_I}|^2 + \frac{p-7}{4} |F_5^{(n)_I}|^2 ) \right) \Bigg) \nn \ .
\eea
We reexpress the source terms using \eqref{Tterms}, the homogeneous overlap \eqref{equalityT10}, and \eqref{R4T10F} to replace $T_{10}$. This gives
\beq
\sum_I e^{\phi} \left( T_{a_{||_I}}^{a_{||_I}}  - \frac{p-3}{p+1} T_{10} \right) =  (N-1) (3-p+ N_o) \left({\cal R}_4 + e^{2\phi} \sum_{q=0}^6 |F_q|^2 \right) \ . \label{newTterms2}
\eeq
Replacing in \eqref{newR41}, we get
\bea
& \frac{1}{2}{\cal R}_4 \left( -4 + (N-1) (3-p + 2N_o) \right)  \label{newR42}\\
& = (-1)^{p} 2\varepsilon_p e^{\phi} \sum_I *_6 \d ( {\rm vol}_{||_I}\w  F_k^{(0)_I})  - \sum_{I\neq J} {\cal O}_{IJ} + \sum_I (2 {\cal R}_{||_I} + 2 {\cal R}_{||_I}^{\bot_I} - |H^{(2)_I}|^2 -2|H^{(3)_I}|^2 ) \nn\\
& + \left|(-1)^{p} \varepsilon_p *_6 \d\big( \sum_I {\rm vol}_{||_I} \big) - e^{\phi} F_k \right|^2 + \sum_I \left|*_{\bot_I}H|_{\bot_I} + \varepsilon_p e^{\phi} F_{k-2}|_{\bot_I} \right|^2  \nn \\
&  -(N-1) (|H|^2 + e^{2\phi} |F_{k-2}|^2 )  \nn\\
& + e^{2\phi} |F_{k}|^2   - (N-1) (3-p+ N_o) e^{2\phi} (|F_{k-4}|^2 + |F_{k-2}|^2 + |F_{k}|^2 ) - \sum_I e^{2\phi} \big(  |F_{k}|^2 - |F_{k}|_{\bot_I}|^2 \big) \nn\\
& - e^{2\phi} \bigg( ((N-1)(4-p+ N_o) -2) |F_{k+2}|^2 + ((N-1) (2(6-p)+N_o) + 5-p) |F_{k+4}|^2 \nn\\
& \phantom{- e^{2\phi} \bigg(}+ (N-1) N_o |F_{k+6}|^2 \bigg) - \sum_I \frac{1}{2} e^{2\phi} \bigg( |(*_6 F_{5})|_{\bot_I}|^2 - |F_{5}|_{\bot_I}|^2 \bigg) \nn\\
& - e^{2\phi} \sum_I \sum_{n=2}^{p-3} (n-1) \left( |F_k^{(n)_I}|^2 + |F_{k+2}^{(n)_I}|^2 + \frac{p-6}{2} |F_{k+4}^{(n)_I}|^2 + \frac{p-7}{4} |F_5^{(n)_I}|^2 \right) \ . \nn
\eea
We notice that for $2 \leq n \leq p-3$, $F_k^{(n)_I}$ can only be $F_k^{(2)_I}$. Using this and replacing $|H|^2$ with \eqref{4dtracefinalFk}, we finally obtain
\bea
\boxed{\mbox{Result:}}\quad & -((N-1) (p-3 - N_o) +2) {\cal R}_4  \nn\\
& = (-1)^{p} 2\varepsilon_p e^{\phi} \sum_I *_6 \d ( {\rm vol}_{||_I}\w  F_k^{(0)_I}) + \sum_I e^{2\phi} |F_{k}^{(0)_I}|^2  \label{newR43}\\
& + \left|(-1)^{p} \varepsilon_p *_6 \d\big( \sum_I {\rm vol}_{||_I} \big) - e^{\phi} F_k \right|^2 + \sum_I \left|*_{\bot_I}H|_{\bot_I} + \varepsilon_p e^{\phi} F_{k-2}|_{\bot_I} \right|^2  \nn \\
& + (N-1) e^{2\phi} |F_{k}|^2 - e^{2\phi} \sum_I |F_k^{(2)_I}|^2 - \sum_{I\neq J} {\cal O}_{IJ} + \sum_I (2 {\cal R}_{||_I} + 2 {\cal R}_{||_I}^{\bot_I} - |H^{(2)_I}|^2 -2|H^{(3)_I}|^2 ) \nn\\
& + (N-1) (p-5 - N_o) e^{2\phi} (|F_{k-4}|^2 + |F_{k-2}|^2 + |F_{k}|^2 )  \nn\\
& + e^{2\phi} \bigg( ((N-1)(p-3 - N_o) +2) |F_{k+2}|^2 + ((N-1) (2(p-5) -N_o) + p-5) |F_{k+4}|^2 \nn\\
& \phantom{- e^{2\phi} \bigg(}+ (N-1) (3-N_o) |F_{k+6}|^2 \bigg) - \sum_I \frac{1}{2} e^{2\phi} \bigg( |(*_6 F_{5})|_{\bot_I}|^2 - |F_{5}|_{\bot_I}|^2 \bigg) \nn\\
& - e^{2\phi} \sum_I \sum_{n=2}^{p-3} (n-1) \left( |F_{k+2}^{(n)_I}|^2 + \frac{p-6}{2} |F_{k+4}^{(n)_I}|^2 + \frac{p-7}{4} |F_5^{(n)_I}|^2 \right) \ . \nn
\eea

This expression \eqref{newR43} should be compared to \eqref{finalhomo}. The only differences are in the $F_k$ terms and the ${\cal O}_{IJ}$ terms; all other flux terms are the same as before. As for \eqref{finalhomo}, in the case $0 \leq N_o \leq p-5$ with $p \geq 5$, all terms in the right-hand side of \eqref{newR43} are of definite sign (upon integration), namely positive, except the line with curvature terms. We deduce the following requirement for de Sitter:
\bea
& \hspace{0.5in} \mbox{For $p\geq 5$ with $0\leq N_o \leq p-5$, having a de Sitter solution requires} \label{requireNonew}\\
& \hspace{-0.3in} \int_6 {\rm vol}_6 \left( (N-1) e^{2\phi} |F_{k}|^2 - e^{2\phi} \sum_I |F_k^{(2)_I}|^2 - \sum_{I\neq J} {\cal O}_{IJ} + \sum_I (2 {\cal R}_{||_I} + 2 {\cal R}_{||_I}^{\bot_I} - |H^{(2)_I}|^2 -2|H^{(3)_I}|^2 ) \right) < 0 \ . \nn
\eea
In practice however, \eqref{requireNonew} does not seem very useful. As argued in Section \ref{sec:foreword}, the expression \eqref{newR43} is rather interesting for Minkowski solutions. Still, the quantity entering \eqref{requireNonew} is the one that should vanish, by analogy to the case of parallel sources \cite{Andriot:2016ufg}. In contrast to the latter though, having it to vanish does not appear as a geometric assumption, due to $F_k$ and $H$. But we can proceed differently: we may first assume the ansatz \eqref{calibflux} that we repeat here
\beq
F_{k} = (-1)^{p} \varepsilon_p e^{-\phi} *_6 \d\big( \sum_J {\rm vol}_{||_J} \big) \ ,\label{calibfluxagain}
\eeq
justified by the calibration of sources. Then, using this and \eqref{squaredvol}, the quantity entering \eqref{requireNonew} becomes a purely geometric quantity, up to $F_k^{(2)_I}$ and the $H$ components. Assuming it to vanish can be viewed in part as a geometric condition on $\mmm$:
\bea
&(N-1) e^{2\phi} |F_{k}|^2 - \sum_{I\neq J} {\cal O}_{IJ} + 2 \sum_I ({\cal R}_{||_I} +   {\cal R}_{||_I}^{\bot_I}) - \sum_I ( e^{2\phi} |F_k^{(2)_I}|^2 + |H^{(2)_I}|^2 +2|H^{(3)_I}|^2 ) \nn\\
=\ & (N-2) \sum_{I\neq J} {\cal O}_{IJ} + (N-1) \sum_I \sum_{a_{||_I}} | ( \d e^{a_{||_I}})|_{\bot_I} |^2 + 2 \sum_I ({\cal R}_{||_I} +   {\cal R}_{||_I}^{\bot_I}) \nn\\
& \phantom{(N-2) \sum_{I\neq J} {\cal O}_{IJ}} \qquad \qquad \qquad - \sum_I ( e^{2\phi} |F_k^{(2)_I}|^2 + |H^{(2)_I}|^2 +2|H^{(3)_I}|^2 ) = 0 \label{geocondnew} \ .
\eea
Interestingly, this seemingly intricate condition simplifies in the case $N=2$; it is worth noticing that the list of examples given in Section \ref{sec:foreword} all have $N=2$. We now come back to finding Minkowski solutions.

\subsection{Towards solutions}\label{sec:Minksolfinal}

The reasoning presented in Section \ref{sec:foreword} consists in making an assumption and deduce from the ${\cal R}_4$ expression an ansatz for the fields, that leads to Minkowski solutions. There are here two options: either one assumes the first line of \eqref{geocondnew} to vanish, and deduces from \eqref{newR43} an ansatz on the fields that includes the expression \eqref{calibfluxagain} for $F_k$; or one assumes this \eqref{calibfluxagain}, justified by the calibration of sources, and the last lines of \eqref{geocondnew}, closer to a geometric condition. In both cases, proceeding as in Section \ref{sec:foreword} (see in particular below \eqref{ansatz0}), one is led to consider $N_o = p-5 \geq 0$ with $N>1$,\footnote{\label{foot:Minkp=4}We have considered \eqref{newR43} in the case $p=4$ with $N_o=0$ and $N>1$, but have not obtained better results than before: we have not found solutions, nor disproved their existence, either for Minkowski or de Sitter.} and the following ansatz is obtained
\bea
& F_k^{(0)_I}= 0 \ \forall I \ ,\ F_{k} = (-1)^{p} \varepsilon_p e^{-\phi} *_6 \d\big( \sum_I {\rm vol}_{||_I} \big) \label{ansatz} \\
& F_{k-2}|_{\bot_I} = - \varepsilon_p e^{-\phi} *_{\bot_I}H|_{\bot_I} \nn\\
& F_{k-4}=F_{k+2}=F_{k+4}= F_{k+6} = 0 \ . \nn
\eea
Let us briefly comment on how this solves the problem encountered in Section \ref{sec:foreword}. We deduce from \eqref{dFnew} and the above ansatz that
\beq
-2\varepsilon_p e^{\phi} \sum_I (\d F_k)_{\bot_I} = - 2 e^{2\phi} |F_k|^2 \ .\label{dFFnew}
\eeq
In Section \ref{sec:foreword}, the problem was raised when comparing to \eqref{firstsquaregen}. With the field ansatz, that equation now gets reduced for Minkowski to
\beq
|H|^2 - \sum_I |H|_{\bot_I}|^2 + e^{2\phi} (  |F_{k-2}|^2 - \sum_I |F_{k-2}|_{\bot_I}|^2 ) = 0 \ .\label{newbla}
\eeq
At least in the case where $H=F_{k-2}=0$, this condition can be satisfied, so it is not problematic anymore.\footnote{One can obtain here the analogue of a no-go theorem for de Sitter solutions obtained in \cite{Andriot:2016xvq} for Minkowski-type calibrations, however only when $H|_{\bot_I}$ or $F_{k-2}|_{\bot_I}$ vanishes. Starting with \eqref{calibfluxagain} for $F_k$, one deduces that \eqref{dFFnew} holds upon integration. Then, integrating \eqref{firstsquaregen}, and assuming $H|_{\bot_I}$ or $F_{k-2}|_{\bot_I}$ to vanish, one concludes that a de Sitter solution cannot be obtained.} In addition, the ansatz now stands the comparison to the known examples; in particular, the curvature terms do not need to vanish by themselves anymore, but rather satisfy \eqref{geocondnew}.

We can actually learn more on the fluxes $H$ and $F_{k-2}$: from \eqref{4dtracefinalFk}, we deduce
\beq
|H|^2 = e^{2\phi} |F_{k-2}|^2 \ ,
\eeq
while we already know from \eqref{ansatz} that $|H|_{\bot_I}|^2 = e^{2\phi} | F_{k-2}|_{\bot_I}|^2$. We deduce with \eqref{newbla} that
\beq
|H|^2 - \sum_I |H|_{\bot_I}|^2 = e^{2\phi} (  |F_{k-2}|^2 - \sum_I |F_{k-2}|_{\bot_I}|^2 ) = 0 \ .
\eeq
This is difficult to satisfy, as discussed below \eqref{firstsquaregen}. In addition, among $p=5,6,7$ on which we focus here with $N>1$, one can verify that only $p=5$ may have these fluxes non-zero. Therefore, we rather consider in the following the case where they vanish. Another take on $H$ and $F_{k-2}$ contributions is proposed in Appendix \ref{ap:HF}: we rewrite the square of their BPS-like condition in a similar fashion to that of $F_k$, by bringing the sum inside the square. This may offer another way to get them non-zero.\\

We now restart completely the reasoning, assuming \eqref{calibfluxagain} for $F_k$ and $F_{k-2}|_{\bot_I} = H|_{\bot_I} = 0$. Interestingly, using only \eqref{dFFnew} (that holds upon integration without assuming $F_{k}^{(0)_I}=0$) and \eqref{firstsquaregen}, one shows that any other flux than $F_k$ vanishes. This is obtained without assuming the condition \eqref{geocondnew}; rather, because of \eqref{newR43}, the latter would have to hold provided $F_{k}^{(0)_I}=0$. This is an alternative way to reach the same field ansatz, with the only non-zero flux $F_k$ given in \eqref{calibfluxagain}.

We now try to prove that this is automatically a solution, as in \cite{Andriot:2016ufg} for parallel sources. The internal Einstein equations will make it too involved, so we only sketch the first steps. The flux e.o.m. are all satisfied. Indeed, in all flux equations but the one of $F_k$, the latter appears times another flux which vanishes. In the $F_k$ equation, the only non-trivial term is $\d *_6 F_k$, which vanishes thanks to the expression \eqref{calibfluxagain}. We turn to the flux BI: the only non-trivial one is that of $F_k$, that we assume to hold as in \cite{Andriot:2016ufg}. It is given by $(\d F_k)_{\bot_I} =  \varepsilon_p\, \tfrac{T_{10}^I}{p+1} $, from which we deduce with \eqref{dFFnew} that
\beq
e^{\phi} \frac{T_{10}}{p+1} = \varepsilon_p e^{\phi} \sum_I (\d F_k)_{\bot_I} =  e^{2\phi} |F_k|^2 \ .\label{BIsolv}
\eeq
Turning to the other equations, we consider the combination of the dilaton e.o.m.~with the ten-dimensional Einstein trace, given in \eqref{10dtracesansR10}: here, it becomes
\beq
(p-3) e^{\phi} \frac{{T}_{10}}{p+1}  - e^{2\phi} (p-3) |F_k|^2 = 0\ .
\eeq
This is satisfied thanks to \eqref{BIsolv}. We are only left with the Einstein equation.

As we consider $p=5,6,7$ and only $F_k$, the only flux is one among $F_{1,2,3}$. We write the Einstein equation accordingly, with constant dilaton, from \cite{Andriot:2016xvq}
\bea
{\cal R}_{MN}-\frac{g_{MN}}{2} {\cal R}_{10} & = \frac{e^{2\phi}}{2} F_{2\ MP}F_{2\ N}^{\ \ \ \ P} + \frac{e^{\phi}}{2}T_{MN} -\frac{g_{MN}}{4} e^{2\phi} |F_2|^2 \ ,  \\
{\cal R}_{MN}-\frac{g_{MN}}{2} {\cal R}_{10} & = \frac{e^{2\phi}}{2}\left(F_{1\ M}F_{1\ N} +\frac{1}{2!} F_{3\ MPQ}F_{3\ N}^{\ \ \ \ PQ} \right) + \frac{e^{\phi}}{2}T_{MN} -\frac{g_{MN}}{4} e^{2\phi}(|F_1|^2 + |F_3|^2 )\nn \ ,
\eea
where in type IIB, one should pick only one of the two fluxes. The ten-dimensional trace becomes
\beq
4 {\cal R}_{10}  + \frac{e^{\phi}}{2} {T}_{10} - \frac{e^{2\phi}}{2} (p-3) |F_k|^2 = 0\ . \label{10dtracesimple}
\eeq
The trace-inversed Einstein equations are thus
\bea
{\cal R}_{MN} & = -\frac{g_{MN}}{16} e^{\phi}T_{10} + \frac{e^{2\phi}}{2} F_{2\ MP}F_{2\ N}^{\ \ \ \ P} + \frac{e^{\phi}}{2}T_{MN} -\frac{g_{MN}}{16} e^{2\phi} |F_2|^2 \label{Einstsimpleti}\ ,  \\
{\cal R}_{MN} & = -\frac{g_{MN}}{16} e^{\phi}T_{10} + \frac{e^{2\phi}}{2}\left(F_{1\ M}F_{1\ N} +\frac{1}{2!} F_{3\ MPQ}F_{3\ N}^{\ \ \ \ PQ} \right) + \frac{e^{\phi}}{2}T_{MN} -\frac{g_{MN}}{8} e^{2\phi} |F_3|^2 \nn \ .
\eea
The four-dimensional trace-inversed equations can be written as
\beq
0 = -\frac{\eta_{\alpha\beta}}{16} e^{\phi}T_{10} + \frac{e^{\phi}}{2}T_{\alpha\beta} -\frac{\eta_{\alpha\beta}}{16} e^{2\phi} |F_2|^2 \ ,\quad 0 = -\frac{\eta_{\alpha\beta}}{16} e^{\phi}T_{10} + \frac{e^{\phi}}{2}T_{\alpha\beta} -\frac{\eta_{\alpha\beta}}{8} e^{2\phi} |F_3|^2 \ .
\eeq
Using $T_{\alpha\beta} = \eta_{\alpha\beta} \tfrac{T_{10}}{p+1}$ and \eqref{BIsolv}, these equations are solved for $p=5,6,7$ (for $p=7$, we recall that only $F_1$ is non-zero).

We are left with the internal Einstein equation. We consider it in the flat basis. We start with the energy momentum tensor: along internal flat directions, it is given as follows, thanks to \eqref{T1} and \eqref{T2}
\beq
T_{ab}= \sum_J \delta_a^{a_{||_J}} \delta_b^{b_{||_J}} \, \delta_{a_{||_J}b_{||_J}} \frac{T_{10}^J}{p+1} \ .
\eeq
To illustrate the difficulties, we specialize to $N=2$ where the two sets of sources are denoted $I$ and $J$. Assuming for simplicity a global basis, the internal space gets split into four sets of directions:\\
- $i$: $p-3 -N_o = 2$ directions $|| I, \bot J$,\\
- $ii$: $N_o =p-5$ directions $|| I, || J$,\\
- $iii$: $p-3 -N_o = 2$ directions $|| J, \bot I$,\\
- $iv$: $6-(N_o + 2 + 2) = 7-p $ directions $\bot I, \bot J$.\\
The energy momentum tensor becomes on each of those
\beq
T_{ab}|_i = \delta_{a_{||_I}b_{||_I}} \frac{T_{10}^I}{p+1} \ ,\ T_{ab}|_{ii} = \delta_{a_{||_I}b_{||_I}} \frac{T_{10}}{p+1}\ , \ T_{ab}|_{iii} = \delta_{a_{||_J}b_{||_J}} \frac{T_{10}^J}{p+1}\ , \ T_{ab}|_{iv} = 0 \ .
\eeq
The internal trace-inversed Einstein equation should be considered on each of those four sets. Starting with $ii$, we obtain an analogous cancelation to that of the four-dimensional components. We are then left with
\beq
ii:\ {\cal R}_{MN}  = \frac{e^{2\phi}}{2}\left(F_{1\ M}F_{1\ N} +  F_{2\ MP}F_{2\ N}^{\ \ \ \ P}  +\frac{1}{2!} F_{3\ MPQ}F_{3\ N}^{\ \ \ \ PQ} \right) \ ,
\eeq
where one only picks one flux, according to the source, and the equation should be projected with vielbeins along $ii$. Computing the flux contribution, given the expression \eqref{BPSlikemultiple} for $F_k$, requires to compare $*_{\bot_J} ( \d e^{b_{||_J}})|_{\bot_J}$ and $*_{\bot_I} ( \d e^{b_{||_I}})|_{\bot_I}$, by decomposing on the four above sets. This is complicated, in particular due to the Hodge star. Along $i$, the equation gets even more involved. In addition, the Ricci tensor is difficult to treat, despite having an expression for it: it probably requires geometrical constraints. Solving these internal Einstein equations is thus difficult in full generality, even though it could be done in concrete examples. This prevents us for now from obtaining a class of Minkowski solutions with intersecting sources, even though we have a well-motivated ansatz and several known examples.

\section{Sources of multiple sizes: no-go theorem for $p=3\, \&\, 7$}\label{sec:multiplesizes}

In this section, we allow for sources of multiple sizes $p$ and study the possibility of getting de Sitter solutions. This work is placed at the end of the paper to avoid confusions, because generalizing to multiple sizes leads to changes in the equations and requires to refine the notations used so far. In particular, we need to pay attention to the traces of the sources energy momentum tensor. While $T_{MN}$ remains defined formally as before in \eqref{defTmnmain}, as well as the overall trace $T_{10}= g^{MN} T_{MN} = \eta^{AB} T_{AB}$, the sum on sources in the different components of $T_{AB}$ now has to be split into a further sum over the different $p$ values. We then define
\beq
T_{10}= \sum_p T_{10}^p \ ,\quad T_{10}^p = - 2 \kappa_{10}^2 T_p (p+1) \sum_{p-{\rm sources}} c_p \left(*_{\bot} \delta^{\bot}_{9-p}\right) \ ,\label{T10p}
\eeq
referring to Appendix \ref{ap:Tmn} for the notations. One can further decompose the last sum into a sum over $I$, as e.g.~in \eqref{traceap}, thus introducing a $T_{10}^{p\,I}$. Going back to previous notations in case of a single size $p$ simply amounts to drop the upper labels ${}^p$ in the above. The BI for the RR fluxes are now written as in \eqref{BI3}, replacing $T_{10}^{I}$ with $T_{10}^{p\,I}$. Indeed, in our setting, we do not consider higher order corrections (see e.g.~\cite{Dasgupta:1999ss}) nor a world-volume $b$-field or ${\cal F}$, so the BI are only sensitive to sources of a single size. Complications appear with the dilaton e.o.m.: one can now verify that
\beq
\frac{1}{\sqrt{|g_{10}|}} \sum_{{\rm sources}} \frac{\delta S_{DBI}}{\delta \phi}=- \frac{e^{- \phi}}{2 \kappa_{10}^2} \sum_p \frac{T_{10}^p}{p+1} \ .
\eeq
That quantity will be important so we denote it as follows
\beq
T_{10p}= \sum_p \frac{T_{10}^p}{p+1} \ .
\eeq
The difference with the previous $\tfrac{T_{10}}{p+1}$ in \eqref{dilcontrib} for a single size $p$ is what makes computations more involved. The dilaton e.o.m., the ten-dimensional Einstein trace, and the four-dimensional one, now become
\bea
& \hspace{-0.1in} 2 {\cal R}_{10} + e^{\phi} T_{10p} -|H|^2 + 8(\Delta \phi - |\del \phi|^2 ) = 0 \ ,\label{dileom2multiple}\\
& \hspace{-0.1in} 4 {\cal R}_{10}  + \frac{e^{\phi}}{2} {T}_{10} - |H|^2 - \frac{e^{2\phi}}{2} \sum_{q=0}^6 (5-q) |F_q|^2 -20 |\del \phi|^2 + 18 \Delta \phi = 0\ , \label{10dtracemultiple}\\
& \hspace{-0.1in} {\cal R}_4 - 2{\cal R}_{10} - 2 e^{\phi} T_{10p} + |H|^2 + e^{2\phi} \sum_{q=0}^6 |F_q|^2  +2 (\nabla\del \phi)_4 + 8 |\del \phi|^2 - 8 \Delta \phi = 0 \ ,\label{4dtracemultiple}
\eea
with even/odd RR fluxes in IIA/IIB, and $g^{MN} T_{MN=\mu\nu} = 4 T_{10p}$.

From now on, we consider the dilaton to be constant. We first proceed as in Section \ref{sec:p=7,8}: we eliminate (part of) the sources contributions in the two Einstein traces. Combining the four-dimensional trace and the dilaton e.o.m. gives
\beq
{\cal R}_4 + 2{\cal R}_{10} - |H|^2 + e^{2\phi} \sum_{q=0}^6 |F_q|^2 = 0 \ . \label{multi4d}
\eeq
For the other trace, we rewrite the dilaton e.o.m. as follows introducing a parameter $p_0\geq 3$:
\beq
e^{\phi} T_{10} = -(p_0 +1) ( 2 {\cal R}_{10} -|H|^2) + e^{\phi} ( T_{10} - (p_0 +1)  T_{10p}) \ .
\eeq
Tuning this $p_0$ allows to erase the $p_0$-source contribution, as can be seen in the last term above. Combining with the ten-dimensional trace, one obtains
\beq
2(3-p_0) {\cal R}_{10}  +  (p_0-1) |H|^2 - e^{2\phi} \sum_{q=0}^6 (5-q) |F_q|^2 +  e^{\phi} ( T_{10} - (p_0 +1)  T_{10p}) = 0 \ .\label{multi10d}
\eeq
Multiplying \eqref{multi4d} by $(3-p_0)$, and combining it with \eqref{multi10d}, finally gives
\beq
(p_0-3) {\cal R}_4 = - 2 |H|^2 + e^{2\phi} \sum_{q=0}^6 (8-p_0 -q) |F_q|^2  + e^{\phi} ( (p_0 +1)  T_{10p}- T_{10}) \ . \label{multifinal1}
\eeq
This is the analogue of \eqref{4dfinalIIA} and \eqref{4dfinalIIB}.

Before studying further \eqref{multifinal1}, let us look at the result rather obtained by proceeding as in Section \ref{sec:derivgen}. Combining the four-dimensional trace and dilaton e.o.m. to eliminate ${\cal R}_{10}$ gives
\beq
{\cal R}_4 =  e^{\phi} T_{10p} - e^{2\phi} \sum_{q=0}^6 |F_q|^2 \label{multiMald} \ .
\eeq
One deduces the requirement for de Sitter solutions
\beq
T_{10p} > 0 \ .
\eeq
Interestingly, as for the previous distinction between $T_{10}$ and $T_{10}^I$, discussed in footnote \ref{foot:signT10I}, it is here unclear that each $T_{10}^p$ needs to be positive or zero. If however they are, one shows the further requirement of having $T_{10}>0$. If we now combine the ten-dimensional trace and dilaton e.o.m. to eliminate ${\cal R}_{10}$, we obtain
\beq
2 |H|^2 - e^{2\phi} \sum_{q=0}^6 (5 -q) |F_q|^2  + e^{\phi} ( T_{10}- 4 T_{10p}) = 0 \label{multi10dother} \ .
\eeq
Multiplying \eqref{multiMald} by a parameter $-\alpha$ and adding it to \eqref{multi10dother}, we get
\beq
-\alpha {\cal R}_4 =  2 |H|^2 - e^{2\phi} \sum_{q=0}^6 (5 - \alpha -q) |F_q|^2  + e^{\phi} ( T_{10}- (\alpha + 4) T_{10p}) \label{multifinal2} \ .
\eeq
For $\alpha=p_0+1$, we obtain the analogue of what is done in Section \ref{sec:derivgen}, while for $\alpha=p_0-3$, we recover \eqref{multifinal1}.\\

We now focus on \eqref{multifinal1} and choose $p_0=7$. In type IIB with $p=3,5,7$ sources, this gives
\beq
4 {\cal R}_4 = - 2 |H|^2 - e^{2\phi} \sum_{q=1}^5 (q-1) |F_q|^2  + e^{\phi} ( T_{10}^3+ \frac{1}{3} T_{10}^5) \ . \label{multibla71}
\eeq
Without other source than $p=7$, this would have reproduced \eqref{4dfinalIIB}. The interest of $p_0=7$ is precisely to drop the $p=7$ source contribution. We now consider the $p=3$ sources: for those, we use the same reasoning as in Section \ref{sec:derivgen}, namely using the BI. As mentioned below \eqref{T10p}, we use the BI \eqref{BI3}, and because there is only $N=1$ set for $p=3$ sources, one has (with $\varepsilon_3 = -1$)
\beq
e^{\phi} \frac{T_{10}^3}{2} = -2 e^{\phi} (\d F_5)_6 + 2 e^{\phi} (H \w F_3)_6 =  -2 e^{\phi} (\d F_5)_6 + |H|^2 + e^{2\phi} |F_3|^2 - \left|*_6 H - e^{\phi}  F_3 \right|^2 \ , \nn
\eeq
where $(\d F_5)_6 = *_6 \d F_5$, etc. Equation \eqref{multibla71} becomes
\beq
{\cal R}_4 = - e^{\phi} (\d F_5)_6  - e^{2\phi} |F_5|^2 - \frac{1}{2} \left|*_6 H - e^{\phi}  F_3 \right|^2  + \frac{e^{\phi}}{12} T_{10}^5 \ . \label{multi7final1}
\eeq
Integrating over $\mmm$, one obtains
\beq
{\cal R}_4 \int {\rm vol}_6 =  -  \int {\rm vol}_6 \left(  e^{2\phi} |F_5|^2 + \frac{1}{2} \left|*_6 H - e^{\phi}  F_3 \right|^2  - \frac{e^{\phi}}{12} T_{10}^5 \right) \ . \label{multi7finalint}
\eeq
We conclude, in our setting:
\bea
&\boxed{\mbox{Result:}}\nn \\
& \mbox{There is no classical de Sitter solution for any combination of $D_3/O_3$ and $D_7/O_7$.} \label{result37}\\
& \mbox{The same holds having in addition $D_5/O_5$, as long as $T_{10}^5 \leq 0$, i.e.~with more $D_5$ than $O_5$.}\nn
\eea
To reach this result, we have combined the techniques allowing us to prove the absence of solution for $p=3$ and $p=7$ separately. We do not manage to prove other strong results in IIB. It would be interesting to study further the particular case of a group manifold: there, the presence of an $O_3$ would force all structure constants to vanish. This could forbid any solution despite the presence of other sources.\\

We turn to type IIA with $p=4,6,8$ sources. Choosing $p_0=8$ in \eqref{multifinal1}, we get
\beq
5 {\cal R}_4 = - 2 |H|^2 - e^{2\phi} \sum_{q=0}^6 q |F_q|^2  + e^{\phi} ( \frac{4}{5} T_{10}^4 + \frac{2}{7} T_{10}^6) \ .
\eeq
Using the BI, one has (with $\varepsilon_4 = -1$)
\beq
e^{\phi} \frac{2}{5} T_{10}^{4I} = - 2 e^{\phi} (\d F_4)_{\bot_I} + 2 e^{\phi} (H \w F_{2})_{\bot_I} =  - 2 e^{\phi} (\d F_4)_{\bot_I} + |H|_{\bot_I}|^2 + e^{2\phi} |F_{2}|_{\bot_I}|^2 - \left|*_{\bot_I} H|_{\bot_I} - e^{\phi}  F_{2}|_{\bot_I} \right|^2 \nn
\eeq
where the $I$, $||_I$ and  $\bot_I$ refer to the $p=4$ sources. This gives
\bea
5 {\cal R}_4  = & - 2 e^{\phi} \sum_I (\d F_4)_{\bot_I} - e^{2\phi} ( 6 |F_6|^2 + 4 |F_4|^2) - 2 \sum_I \left|*_{\bot_I} H|_{\bot_I} - e^{\phi}  F_{2}|_{\bot_I} \right|^2 \\
&  - 2 ( |H|^2 - \sum_I |H|_{\bot_I}|^2 ) - e^{2\phi} 2 ( |F_2|^2 - \sum_I |F_{2}|_{\bot_I}|^2 ) + e^{\phi} \frac{2}{7} T_{10}^6 \ . \nn
\eea
We now restrict ourselves to $N=1$ set for each size $p=4,6,8$. We anticipate on the difficulties that could otherwise appear due to $p=4$, as seen in Section \ref{sec:nogo456}. We further restrict, for future convenience, to having the $p=4$ sources inside the $p=6$, themselves inside the $p=8$. Such a parallel configuration may also preserve some supersymmetry in the four-dimensional theory. We then rewrite the above by dropping the label $I$ (since $N=1$) and replacing it with a $4$, to indicate that we refer to the transverse space of the $p=4$, etc.:
\bea
5 {\cal R}_4  = & - 2 e^{\phi} (\d F_4)_{\bot_4} - e^{2\phi} ( 6 |F_6|^2 + 4 |F_4|^2) - 2 \left|*_{\bot_4} H|_{\bot_4} - e^{\phi}  F_{2}|_{\bot_4} \right|^2 \\
&  - 2 ( |H|^2 - |H|_{\bot_4}|^2 ) - e^{2\phi} 2 ( |F_2|^2 -  |F_{2}|_{\bot_4}|^2 ) + e^{\phi} \frac{2}{7} T_{10}^6 \ . \nn
\eea
We now proceed as usual with
\bea
2  e^{\phi} (\d F_4)_{\bot_4} = &  2 e^{\phi} (\d F_4^{(0)_4})_{\bot_4} +  \left| *_{\bot_4}( \d e^{a_{||_4}})|_{\bot_4} + e^{\phi}\, \iota_{a_{||_4}} F_4^{(1)_4} \right|^2  \\
& \phantom{2 e^{\phi} (\d F_k^{(0)_I})_{\bot_I}}\ - e^{2\phi} |F_4^{(1)_4}|^2 -  |(\d e^{a_{||_4}})|_{\bot_4}|^2  \ , \nn
\eea
where the $\sum_{a_{||_I}}$ is dropped because there is only one parallel direction for $p=4$. We then compute the trace of the internal Einstein equation along that direction. The result is combined with the four-dimensional trace, where we now have $T_{10p}$ instead of $T_{10}/(p+1)$. This is nicely compensated by $\eta^{AB}T_{AB=a_{||_J}b_{||_J}}$, thanks to the overlap of $p=6,8$ sources with the single direction of the $p=4$ sources. This trace becomes
\bea
{\cal R}_{6||_4} & = \frac{1}{4} \left({\cal R}_4 +  2e^{2\phi} |F_{6}|^2  \right) \\
& + \frac{1}{2} \left(|H|^2 - |H|_{\bot_4}|^2 + e^{2\phi} (  |F_{2}|^2 - |F_{2}|_{\bot_4}|^2 + |F_{4}|^2 - |F_{4}|_{\bot_4}|^2 \right) \nn \\
& = {\cal R}_{||_4} +  {\cal R}_{||_4}^{\bot_4} + \frac{1}{2}  |(\d e^{a_{||_4}})|_{\bot_4}|^2 \ .
\eea
We deduce
\beq
\mbox{In this setting, a de Sitter solution requires $- \frac{1}{2}  |(\d e^{a_{||_4}})|_{\bot_4}|^2 < {\cal R}_{||_4} +  {\cal R}_{||_4}^{\bot_4} $} \ . \label{requiremulti4}
\eeq
As pointed-out in \cite{Andriot:2016xvq}, $f^{a_{||_4}}{}_{b_{||_4}c_{||_4}}=0$ because there is only one internal parallel direction, so ${\cal R}_{||_4} =0$. Combining these results and using $|F_4|^2 = |F_4^{(0)_4}|^2 +|F_4^{(1)_4}|^2 $, we obtain
\bea
\frac{9}{2} {\cal R}_4  = & - 2 e^{\phi} (\d F_4^{(0)_4})_{\bot_4} -  2 e^{2\phi} |F_4^{(0)_4}|^2 - 2 \left|*_{\bot_4} H|_{\bot_4} - e^{\phi}  F_{2}|_{\bot_4} \right|^2 - \left| *_{\bot_4}( \d e^{a_{||_4}})|_{\bot_4} + e^{\phi}\, \iota_{a_{||_4}} F_4^{(1)_4} \right|^2 \nn\\
&  - e^{2\phi} ( 5 |F_6|^2 + 2 |F_4|^2) -  ( |H|^2 - |H|_{\bot_4}|^2 ) - e^{2\phi}  ( |F_2|^2 -  |F_{2}|_{\bot_4}|^2 )  \\
&  - 2 {\cal R}_{||_4}^{\bot_4} + e^{\phi} \frac{2}{7} T_{10}^6 \ . \nn
\eea
Proceeding as in \eqref{inttrick} for the integration, we deduce
\bea
\frac{9}{2} {\cal R}_4 \int {\rm vol}_6  =   - \int {\rm vol}_6  & \Big( 2 e^{2\phi} |F_4^{(0)_4}|^2 + 2 \left|*_{\bot_4} H|_{\bot_4} - e^{\phi}  F_{2}|_{\bot_4} \right|^2 + \left| *_{\bot_4}( \d e^{a_{||_4}})|_{\bot_4} + e^{\phi}\, \iota_{a_{||_4}} F_4^{(1)_4} \right|^2 \nn\\
&  + e^{2\phi} ( 5 |F_6|^2 + 2 |F_4|^2) +  ( |H|^2 - |H|_{\bot_4}|^2 ) + e^{2\phi}  ( |F_2|^2 -  |F_{2}|_{\bot_4}|^2 ) \nn \\
&  + 2 {\cal R}_{||_4}^{\bot_4} - e^{\phi} \frac{2}{7} T_{10}^6 \Big) \ .
\eea
We conclude
\bea
& \mbox{There is no classical de Sitter solution for a combination of parallel ($N=1$) sets of} \label{result48}\\
& \mbox{$D_4/O_4$, $D_6/O_6$ and $D_8/O_8$, i.e.~included into each other, if} \nn\\
& \mbox{${\cal R}_{||_4}^{\bot_4} \geq 0$ and $T_{10}^6 \leq 0$ (i.e.~with more contributions from $D_6$ than $O_6$).}\nn
\eea
In particular, for the supersymmetric system of parallel $D_4/O_4$ and $D_8/O_8$, the constraint is simply on the curvature term ${\cal R}_{||_4}^{\bot_4} $. Combined with \eqref{requiremulti4}, the conditions obtained are exactly the same as for parallel $D_4/O_4$ alone \cite{Andriot:2016xvq}, while allowing here for additional parallel $D_8/O_8$.

\section{Summary of results and outlook}\label{sec:ccl}

In this paper, we have studied the possibility of getting classical de Sitter or Minkowski solutions of ten-dimensional type II supergravities, with intersecting Ramond-Ramond sources, namely $D_p$-branes and orientifold $O_p$-planes. This motivated by the connection of string theory to both cosmological models and particle physics model building, as presented in the Introduction. While only few explicit solutions are known, this work aims at getting a general characterisation for them. In Section \ref{sec:formal}, we have detailed the framework and few assumptions with which we work, and developed a formalism to treat intersecting sources. The method has then been to derive interesting expressions of the four-dimensional space-time Ricci scalar ${\cal R}_4$ in terms of internal fields. For de Sitter, the requirement is then to have ${\cal R}_4 > 0$, which leads to various constraints, while having Minkowski imposes ${\cal R}_4 =0$, which leads to a solution ansatz for the internal fields. This way, we obtained several results, that we now summarize:
\begin{itemize}
  \item There is no classical de Sitter solution with $D_3/O_3$, or with (intersecting) $D_7/O_7$, or any combination of the two. This was shown respectively in \cite{Giddings:2001yu, Blaback:2010sj, Andriot:2016xvq}, \eqref{nogo78} and \eqref{result37}. We recall that this is valid in our framework, which does not include non-perturbative F-theory solutions. This result should be of interest for many stringy inflation models built with such ingredients, typically on a Calabi-Yau manifold, such as the recent \cite{Landete:2017amp, Cicoli:2017shd}. Those models usually include additional ingredients, mostly at the four-dimensional level, and the present result then provides a further motivation to do so.
  \item There is no classical de Sitter solution with (intersecting) $D_8/O_8$. In addition, solutions with parallel $D_4/O_4$ and $D_8/O_8$ (meaning $N=1$ set of $D_4/O_4$, included in $N=1$ set of $D_8/O_8$) are constrained precisely in the same manner as those with only parallel $D_4/O_4$. This was shown respectively in \cite{Andriot:2016xvq}, \eqref{nogo78} and \eqref{result48}.
  \item Classical de Sitter solutions with intersecting $D_5/O_5$ or $D_6/O_6$ get very interesting constraints in the special case of homogeneous overlap \eqref{assumptionoverlap}, with $0\leq N_o \leq p-5$: they are then constrained by a specific combination of curvature terms and $H$-flux components as given in \eqref{requireNo}. These constraints generalize those obtained for parallel sources \cite{Andriot:2016xvq}, and indicate that de Sitter solutions are easier to obtain with intersecting sources (see also below \eqref{require3}). As a corollary \eqref{nogo6group}, there is no classical de Sitter solution with non-overlapping $O_6$ on a group manifold with constant $H$-flux.
  \item Classical Minkowski solutions with intersecting $D_5/O_5$, $D_6/O_6$, or $D_7/O_7$, were studied in the case of homogeneous overlap \eqref{assumptionoverlap}, with $0\leq N_o \leq p-5$. Contrary to the situation of parallel sources \cite{Andriot:2016ufg}, the ${\cal R}_4$ expression relevant to constrain de Sitter solutions \eqref{finalhomo} is here not appropriate. We then derived another ${\cal R}_4$ expression \eqref{newR43} (see also \eqref{newR43new}) from which one motivates an interesting ansatz of solution, especially for the sourced flux $F_{k=8-p}$ \eqref{calibfluxagain}. We were nevertheless unable for now to prove in full generality that this ansatz is a solution. Still, typical features of such Minkowski solutions were understood: for instance, curvature terms would not vanish for intersecting sources, contrary to parallel ones. Therefore, if one wants to move from simple toroidal solutions (e.g.~to stabilize moduli in a model building context), adding fluxes is not enough: a change in the geometry is also required.
  \item Classical de Sitter or Minkowski solutions with intersecting $D_4/O_4$ are both hard to constrain or to find: see \eqref{requirep=4} and \eqref{require2} for a discussion and constraints on de Sitter solutions, and the end of Section \ref{sec:groupmanif} and Footnote \ref{foot:Minkp=4} for further comments on Minkowski solutions.
  \item An outcome of this analysis with intersecting sources is the importance of the information on the sources overlap. We mostly focused on the ``simple'' case of homogeneous overlap \eqref{assumptionoverlap}. Although restrictive, this case turns-out to be realised in almost all examples of known solutions, and is thus very relevant. First, all known Minkowski solutions with intersecting sources on solvmanifolds, except the torus (see the list in Section \ref{sec:foreword}), have $p=5,6$, $N=2$ and $N_o=p-5$. Second, all known classical de Sitter solutions (except the one with $O_5/O_7$ \cite{Caviezel:2009tu}) admit $N=4$ intersecting $O_6$ with $N_o=1$ (see Footnote \ref{foot:dSorbif} for details). Finally, particle physics model building on torus orbifolds also use this configuration of $N=4$ intersecting $O_6$ with $N_o=1$: as recalled in Footnote \ref{foot:dSorbif}, this is the case of the seminal $T^6/\mathbb{Z}_2 \times \mathbb{Z}_2$ orientifold model with intersecting branes \cite{Forste:2000hx, Cvetic:2001tj, Cvetic:2001nr}.\footnote{On top of the orientifold in the $T^6/\mathbb{Z}_2 \times \mathbb{Z}_2$, one typically adds $D_6$ at angles $< \pi/2$ with respect to the $O_6$. Our description may be able to capture that, either by considering more sets for the $D_6$ with $0<N_o<1$, or by projecting the $D_6$ on the orthonormal basis of the $O_6$ and thus including them in the existing sets. An appeal of this model (that can be viewed as $N=4$ $O_6$ with $N_o=1$ on $T^6$) is that some supersymmetry is preserved provided the $D_6$ angles fulfill some conditions \cite{Berkooz:1996km}; this holds without discrete torsion. An extension of this result with discrete torsion has been obtained in \cite{Blumenhagen:2005tn}, and for a $\mathbb{Z}_2 \times \mathbb{Z}_6$ orbifold in \cite{Ecker:2014hma, Ecker:2015vea}. The latter give further constructions of interesting particle physics models, even though the $O_6$ configuration there is less easily described in our framework: the orbifold action generates discrete orbits of $O_6$-planes rather than having them at fixed loci. It would be interesting to study such configurations in more detail.}
  \item The expressions derived for ${\cal R}_4$ are of general interest: \eqref{finalhomo} (see also \eqref{finalint}) to constrain de Sitter solutions, and \eqref{newR43} (see also \eqref{newR43new}) to find Minkowski solutions. We rewrite \eqref{finalhomo} schematically as
      \bea
      & ((N-1)(p-3 -N_o) + 2) {\cal R}_4 \label{finalhomoccl}\\
      & = - \sum_I \left|*_{\bot_I}H|_{\bot_I} + \varepsilon_p e^{\phi} F_{k-2}|_{\bot_I} \right|^2 - \sum_I \sum_{a_{||_I}} \left| *_{\bot_I}( \d e^{a_{||_I}})|_{\bot_I} - \varepsilon_p e^{\phi}\, \iota_{a_{||_I}} F_k^{(1)_I} \right|^2 \nn\\
      & - e^{2\phi} \sum |{\rm flux}|^2 + \del(\dots) \nn\\
      & - (N-1) (p-5-N_o) e^{2\phi}( |F_{k-4}|^2 + |F_{k-2}|^2 + |F_{k}|^2 ) \nn\\
      & + \sum_I ( -2 {\cal R}_{||_I} - 2 {\cal R}_{||_I}^{\bot_I} + |H^{(2)_I}|^2 + 2 |H^{(3)_I}|^2 )  \ , \nn
      \eea
      where the left hand-side coefficient is strictly positive. The second line contains the analogue of BPS-like conditions, reminiscent of supersymmetric solutions. The third line contains a total derivative term, and flux terms that are all $\leq 0$, provided $p=3,4$, or $p\geq 5$ and $0\leq N_o \leq p-5$. The fourth line consists in an interesting term, purely due to the intersection, that points towards the specific value $N_o=p-5$ observed in the known examples. The last line contains the combination of terms that are subject to the constraints for de Sitter solutions. For $p\geq 5$ and $0\leq N_o \leq p-5$, all terms in the right hand-side are negative or zero, except for the specific terms of the last line. This nicely illustrates how much type II supergravities seem reluctant in admitting de Sitter solutions, compared to Minkowski or anti-de Sitter ones.
\end{itemize}

Concerning de Sitter solutions, the next step is to study their stability, as discussed in the Introduction. The approach described in \cite{Danielsson:2012et}, where a scalar potential for three moduli is considered, could be relevant for us. The new existence constraints derived here, combined with the corresponding stability constraints, could lead to the identification of a systematic tachyon, at least in the case of intersecting $D_5/O_5$ or $D_6/O_6$ with $N_o=p-5$. This would explain the instability observed in all examples of \cite{Danielsson:2011au}.

An alternative approach is that of \cite{Junghans:2016uvg, Junghans:2016abx}. There, a tachyon was shown to appear in four dimensions, provided the de Sitter solution is close to a Minkowski no-scale solution. This mechanism was shown to be at work for two known ten-dimensional de Sitter solutions in \cite{Junghans:2016uvg}. The new characterisation of classical de Sitter solutions derived here may help generalizing this result to all known solutions, at least, thus identifying a systematic tachyon.

Another possible outcome is the identification of a (narrow) window in parameter space, where both existence and metastability can be reached. Such a result would help finding an explicit metastable classical de Sitter solution. If this is achieved with intersecting $O_6$ and $N_o=1$, there is a chance to connect to particle physics model building, which would bring its own constraints, and lead to a very narrow framework where all requirements could be satisfied. A setting adapted to both metastable de Sitter solutions and particle physics models would be ideal to construct models describing the end of cosmological inflation, where reheating should occur and lead to matter formation and radiation.

\vspace{0.3in}

\subsection*{Acknowledgements}

I wish to thank J.~Bl{\aa }b\"ack, G.~Honecker, D.~Junghans, M.~Petrini, G.~Shiu, A.~Tomasiello, J.-P.~Uzan, T.~Van Riet, T.~Weigand and M.~Zagermann for very useful discussions.

\newpage

\begin{appendix}

\section{Sources contributions}\label{ap:Tmn}

We present in this appendix various derivations about the sources energy momentum tensor $T_{MN}$ and its trace $T_{10}$. To that end, we use the properties or assumptions on the sources and the internal geometry detailed at the beginning of Section \ref{sec:formal}, and few other definitions given in that section.

For each source, there is a natural definition of the parallel or transverse Hodge stars, for forms defined on either of these subspaces; it is compatible with the six-dimensional Hodge star in flat indices. For instance,
\beq
*_{\bot} ( e^{a_{1\bot}} \w \dots \w e^{a_{i\bot}}) = \frac{1}{(9-p-i)!}\, \delta^{a_{1\bot} b_{1\bot}} \dots \delta^{a_{i\bot} b_{i\bot}} \epsilon_{b_{1\bot} \dots b_{9-p\bot}} e^{b_{i+1\bot}} \w \dots \w e^{b_{9-p\bot}} \ . \label{Hodgebot}
\eeq
We now consider the action of each source: with assumptions of Section \ref{sec:formal}, it is given by the following terms
\beq
S_{DBI} \stackrel{{\rm (here)}}{===} - c_p\, T_p \int e^{-\phi}\, {\rm vol}_{4} \w {\rm vol}_{||} \w \delta^{\bot}_{9-p}  \ , \ \ S_{WZ} \stackrel{{\rm (here)}}{===}  c_p\, \mu_p \int C_{p+1} \w \delta^{\bot}_{9-p}  \ ,\label{WZ10D}
\eeq
where the form ordering is a convention choice, $c_p=1$ for a $D_p$ and $-2^{p-5}$ for an $O_p$, and we refer to \cite{Andriot:2016xvq} for more details. We have used \eqref{worldvolform} and further introduced the $(9-p)$-form $\delta^{\bot}_{9-p}$, to remove the pull-back and promote the integrals to ten-dimensional ones. Given the volume forms relations \eqref{volrel}, we can restrict $\delta^{\bot}_{9-p}$ to be proportional to ${\rm vol}_{\bot}$. It can be written as
\beq
\delta^{\bot}_{9-p} = \left(*_{\bot} \delta^{\bot}_{9-p}\right)\ {\rm vol}_{\bot} \ . \label{formcoef}
\eeq
If the metric was block diagonal, the coefficient would be the inverse of the transverse metric determinant, times a formal delta function $\delta(\bot)$ that localizes the source in the transverse directions; but we do not restrict to such a case here and work more formally. By definition, $\delta^{\bot}_{9-p}$ does not depend on any vielbein nor any metric. In addition, from \eqref{formcoef}, we deduce
\beq
\frac{\delta \left(*_{\bot} \delta^{\bot}_{9-p}\right)}{\delta e^M{}_{a_{||}}} = 0 \ ,\label{bla1}
\eeq
while
\beq
\frac{\delta {\rm vol}_{4} \w {\rm vol}_{||}}{\delta e^M{}_{a_{\bot}}} = 0 \ . \label{bla2}
\eeq
For each source, the energy momentum tensor $T_{MN}$ is defined as
\beq
\mbox{For one source:}\quad \frac{1}{\sqrt{|g_{10}|}} \frac{\delta S_{DBI}}{\delta g^{MN}} = - \frac{e^{-\phi}}{4 \kappa_{10}^2} T_{M N} \ ,
\eeq
while for several sources, one simply adds each contribution as in \eqref{defTmnmain}. Because $S_{WZ}$ is topological, i.e.~does not depend on $g_{MN}$, it does not contribute to the derivation here. For each source, we now rewrite the above with flat indices, for instance $T_{a_{\bot}b_{\bot}} = e^M{}_{a_{\bot}} e^N{}_{b_{\bot}} T_{MN}$. Given that
\bea
\frac{\delta}{\delta g^{MN}} &= \eta_{CD}e^D{}_N \frac{\delta}{\delta e^M{}_C} + \eta_{CD}e^C{}_M \frac{\delta}{\delta e^N{}_D} \ , \\
e^M{}_{a_{\bot}} e^N{}_{b_{\bot}}\frac{\delta}{\delta g^{MN}} &= e^M{}_{a_{\bot}} \delta_{c_{\bot}b_{\bot}} \frac{\delta}{\delta e^M{}_{c_{\bot}}} + \delta_{a_{\bot}d_{\bot}}e^N{}_{b_{\bot}} \frac{\delta}{\delta e^N{}_{d_{\bot}}} \ ,
\eea
we deduce with \eqref{bla2} that
\beq
\mbox{For one source:}\quad T_{a_{\bot}b_{\bot}} = 0 \ .
\eeq
Furthermore, using the above, especially \eqref{bla1}, we compute
\bea
\mbox{For one source:}\quad T_{a_{||}b_{||}} & = \frac{4 \kappa_{10}^2}{\sqrt{|g_{10}|}} c_p T_p\, e^M{}_{a_{||}} e^N{}_{b_{||}} \frac{\delta \int {\rm vol}_{4} \w {\rm vol}_{||} \w \delta^{\bot}_{9-p} }{\delta g^{MN}} \\
& = \frac{4 \kappa_{10}^2}{\sqrt{|g_{10}|}} c_p T_p\, e^M{}_{a_{||}} e^N{}_{b_{||}} \frac{\delta \sqrt{|g_{10}|}  }{\delta g^{MN}} \left(*_{\bot} \delta^{\bot}_{9-p}\right) \nn \\
& = - 2 \kappa_{10}^2 c_p T_p\, \delta_{a_{||}b_{||}} \left(*_{\bot} \delta^{\bot}_{9-p}\right) \ . \nn
\eea
Finally, the trace $T_{10}= g^{MN} T_{M N}$ is computed by decomposing on each set of directions
\beq
\mbox{For one source:}\quad T_{10}= \delta^{\alpha \beta} T_{\alpha \beta} + \delta^{a_{||}b_{||}} T_{a_{||}b_{||}} + \delta^{a_{\bot}b_{\bot}} T_{a_{\bot}b_{\bot}} = - 2 \kappa_{10}^2 c_p T_p (p+1) \left(*_{\bot} \delta^{\bot}_{9-p}\right) \ .\nn
\eeq

We now turn to having several sources and use notations introduced in Section \ref{sec:formal}. In flat indices, the energy momentum tensor $T_{AB}= e^M{}_A e^N{}_B T_{MN}$ becomes
\bea
T_{AB}=& \frac{4 \kappa_{10}^2}{\sqrt{|g_{10}|}} T_p\ e^M{}_A e^N{}_B \sum_{I} \sum_{{\rm sources}\in I}  c_p\, \frac{\delta \int {\rm vol}_{4} \w {\rm vol}_{||_I} \w \delta^{\bot_I}_{9-p} }{\delta g^{MN}} \\
=& \frac{4 \kappa_{10}^2}{\sqrt{|g_{10}|}} T_p \sum_{I} \left(\delta_A^{\alpha} \delta_B^{\beta} e^M{}_{\alpha} e^N{}_{\beta} + \delta_A^{a_{||_I}} \delta_B^{b_{||_I}} e^M{}_{a_{||_I}} e^N{}_{b_{||_I}} + \delta_A^{a_{\bot_I}} \delta_B^{b_{\bot_I}} e^M{}_{a_{\bot_I}} e^N{}_{b_{\bot_I}} \right) \nn\\
 & \quad \times \sum_{{\rm sources}\in I}  c_p\, \frac{\delta \int {\rm vol}_{4} \w {\rm vol}_{||_I} \w \delta^{\bot_I}_{9-p} }{\delta g^{MN}} \ . \nn
\eea
Using previous results for each source, we deduce that $T_{AB}= \delta_A^{\alpha} \delta_B^{\beta}\, T_{\alpha \beta} + \sum_I \delta_A^{a_{||_I}} \delta_B^{b_{||_I}} \, T^I_{a_{||_I}b_{||_I}}$ as in \eqref{T1}, where
\bea
T_{\alpha \beta} &= - 2 \kappa_{10}^2 T_p\, \eta_{\alpha \beta} \sum_{I} \sum_{{\rm sources}\in I} c_p \left(*_{\bot_I} \delta^{\bot_I}_{9-p}\right) \ , \\
T^I_{a_{||_I}b_{||_I}} &= - 2 \kappa_{10}^2 T_p\, \delta_{a_{||_I}b_{||_I}} \sum_{{\rm sources}\in I} c_p \left(*_{\bot_I} \delta^{\bot_I}_{9-p}\right) \ .
\eea
We then obtain the trace
\beq
T_{10} = - 2 \kappa_{10}^2 T_p (p+1) \sum_{{\rm sources}} c_p \left(*_{\bot} \delta^{\bot}_{9-p}\right) = - 2 \kappa_{10}^2 T_p (p+1) \sum_{I} \sum_{{\rm sources}\in I} c_p \left(*_{\bot_I} \delta^{\bot_I}_{9-p}\right) \ .\label{traceap}
\eeq
Introducing
\beq
T_{10}^I =  - 2 \kappa_{10}^2 T_p (p+1)  \sum_{{\rm sources}\in I} c_p \left(*_{\bot_I} \delta^{\bot_I}_{9-p}\right) \ ,\quad T_{10} = \sum_{I} T_{10}^I \ ,
\eeq
one gets, as given in \eqref{T2},
\beq
T_{\alpha \beta}= \eta_{\alpha \beta} \frac{T_{10}}{p+1}  \ , \quad T^I_{a_{||_I}b_{||_I}} = \delta_{a_{||_I}b_{||_I}} \frac{T_{10}^I}{p+1}  \ .\label{T2ap}
\eeq

With the above definitions, one can verify that the contributions to the dilaton e.o.m. are, as given in \eqref{dilcontrib},
\beq
\frac{1}{\sqrt{|g_{10}|}} \sum_{{\rm sources}} \frac{\delta S_{DBI}}{\delta \phi}=- \frac{e^{- \phi}}{2 \kappa_{10}^2} \frac{T_{10}}{p+1} \ .
\eeq

Finally, the fluxes Bianchi identities (BI) as given in Appendix A of \cite{Andriot:2016xvq} remain valid. Given the present assumptions, the BI simplify: the fixed $p$ selects only one (internal form) flux $F_k$ to be sourced, with the following BI
\bea
&\d F_k - H \w F_{k-2} = -\varepsilon_p\, 2 \kappa_{10}^2\,  T_p \!\!\! \sum_{p-{\rm sources}} \!\!\! c_p\, \delta^{\bot}_{9-p}  \label{BI}\\
& \mbox{for}\ 0 \leq k=8-p \leq 5\ ,\ \varepsilon_p=(-1)^{p+1} (-1)^{\left[\frac{9-p}{2} \right]} \ ,\nn
\eea
with $F_{-1} =F_{-2}=0$. The previously defined quantities allow to rewrite the BI as in \eqref{BI2}.

\section{Reformulating the $H$ and $F_{k-2}$ contributions}\label{ap:HF}

In Section \ref{sec:foreword} and \ref{sec:derivcomment}, we analysed and rewrote the $F_k$ contributions to the ${\cal R}_4$ expression: with respect to \eqref{finalhomo}, we moved the sum on $I$ inside the square, towards \eqref{newR43}. In this appendix, we reach a similar result for $H$ and $F_{k-2}$ contributions, bringing us closer, in a sense, to the supersymmetric case. Starting from the BI \eqref{BI3}, we rewrite
\bea
& 2e^{\phi} \varepsilon_p \sum_I (H\w F_{k-2})_{\bot_I} = 2 e^{\phi} \varepsilon_p \sum_I *_{\bot_I} (H\w F_{k-2})|_{\bot_I} \\
& = 2 e^{\phi} \varepsilon_p \sum_I *_6 \left({\rm vol}_{||_I}\w (H\w F_{k-2})|_{\bot_I} \right) = 2 e^{\phi} \varepsilon_p \sum_I *_6 \left({\rm vol}_{||_I}\w H\w F_{k-2} \right) \nn\\
& = e^{\phi} \varepsilon_p *_6 \left(F_{k-2} \w *_6^2 ( \sum_I {\rm vol}_{||_I}\w H )\right) + e^{\phi} \varepsilon_p *_6 \left(*_6 (\sum_I {\rm vol}_{||_I}\w H) \w *_6 F_{k-2} \right) \nn\\
& = \left| *_6 (\sum_I {\rm vol}_{||_I}\w H) + e^{\phi} \varepsilon_p F_{k-2} \right|^2 - |\sum_I {\rm vol}_{||_I}\w H|^2 - e^{2\phi} | F_{k-2}|^2 \nn \ ,
\eea
and one could also replace $*_6 ({\rm vol}_{||_I}\w H) = *_{\bot_I} H|_{\bot_I} $. The gain is to have now the sum inside the square, and to have the full $F_{k-2}$. As for \eqref{squaredvol}, one gets
\bea
& |\sum_I {\rm vol}_{||_I}\w H|^2 = \sum_I |H|_{\bot_I}|^2 + \sum_{I\neq J} {\cal P}_{IJ} \\
& {\cal P}_{IJ} = *_6 \left( {\rm vol}_{||_I}\w H \w *_6({\rm vol}_{||_J}\w H) \right) \ , \nn
\eea
where $H$ could be reduced to $H|_{\bot_I}$ in those expressions. As with ${\cal O}_{IJ}$, the cost of bringing the sum inside the square is to have the double product terms ${\cal P}_{IJ}$. Using these expressions, one trades \eqref{firstsquaregen} for
\bea
{\cal R}_4  = -\frac{2}{p+1} \bigg( & -  2 \varepsilon_p e^{\phi} \sum_I (\d F_k)_{\bot_I}  +  \left|*_6 (\sum_I {\rm vol}_{||_I}\w H) + \varepsilon_p e^{\phi} F_{k-2} \right|^2  \label{firstsquaregennew}\\
& +|H|^2 - \sum_I |H|_{\bot_I}|^2 - \sum_{I\neq J} {\cal P}_{IJ}  \nn\\
& + e^{2\phi} ( 2 |F_{k}|^2 + 3 |F_{k+2}|^2 + 4 |F_{k+4}|^2 + 5 |F_{k+6}|^2 )  \bigg)  \ . \nn
\eea
We then rewrite \eqref{newR41} as
\bea
{\cal R}_4  = -\frac{2}{p+1} \bigg( & (-1)^{p} 2\varepsilon_p e^{\phi} \sum_I *_6 \d ( {\rm vol}_{||_I}\w  F_k^{(0)_I}) - |\d\big( \sum_I {\rm vol}_{||_I} \big) |^2 \label{newR41new}\\
& + \left|(-1)^{p} \varepsilon_p *_6 \d\big( \sum_I {\rm vol}_{||_I} \big) - e^{\phi} F_k \right|^2 + \left|-(-1)^{p}\varepsilon_p *_6 (H \w \sum_I {\rm vol}_{||_I}) + e^{\phi} F_{k-2} \right|^2  \nn \\
& + \sum_I (|H|^2- |H|_{\bot_I}|^2) -(N-1) |H|^2 - \sum_{I\neq J} {\cal P}_{IJ}  \nn\\
& + e^{2\phi} ( |F_{k}|^2 + 3 |F_{k+2}|^2 + 4 |F_{k+4}|^2 + 5 |F_{k+6}|^2 )  \bigg)  \ , \nn
\eea
where the $H$-flux is put forward in the square. This indicates the possible gathering of the squares of BPS-like conditions towards the known combinations $(\d-H\w)\sum_I {\rm vol}_{||_I}$ and $F_k - F_{k-2}$, familiar from supersymmetry. From \eqref{newR41new}, we proceed as in Section \ref{sec:derivcomment} and end-up with
\bea
\boxed{\mbox{Result:}}\quad & -((N-1) (p-3 - N_o) +2) {\cal R}_4  \nn\\
& = (-1)^{p} 2\varepsilon_p e^{\phi} \sum_I *_6 \d ( {\rm vol}_{||_I}\w  F_k^{(0)_I}) + \sum_I e^{2\phi} |F_{k}^{(0)_I}|^2 \label{newR43new}\\
& + \left|(-1)^{p} \varepsilon_p *_6 \d\big( \sum_I {\rm vol}_{||_I} \big) - e^{\phi} F_k \right|^2 + \left|-(-1)^{p}\varepsilon_p *_6 (H \w \sum_I {\rm vol}_{||_I}) + e^{\phi} F_{k-2} \right|^2  \nn \\
& + (N-1) e^{2\phi} |F_{k}|^2 - e^{2\phi} \sum_I |F_k^{(2)_I}|^2 - \sum_{I\neq J} {\cal O}_{IJ} + \sum_I (2 {\cal R}_{||_I} +  2 {\cal R}_{||_I}^{\bot_I}  - |H^{(2)_I}|^2 - 2 |H^{(3)_I}|^2) \nn\\
& + (N-1) e^{2\phi} |F_{k-2}|^2 - e^{2\phi}  \sum_I ( |F_{k-2}|^2 - |F_{k-2}^{(0)_I}|^2 ) - \sum_{I\neq J} {\cal P}_{IJ} \nn\\
& - (N-1) (5-p+ N_o) e^{2\phi} (|F_{k-4}|^2 + |F_{k-2}|^2 + |F_{k}|^2 )  \nn\\
& \hspace{-1.3in} + e^{2\phi} \bigg( ((N-1)(p-3- N_o) +2) |F_{k+2}|^2 + ((N-1) (2(p-5)-N_o) + p-5) |F_{k+4}|^2 + (N-1) (3-N_o) |F_{k+6}|^2 \bigg) \nn\\
& \hspace{-1.3in} - \sum_I \frac{1}{2} e^{2\phi} \bigg( |(*_6 F_{5})|_{\bot_I}|^2 - |F_{5}|_{\bot_I}|^2 \bigg)  - e^{2\phi} \sum_I \sum_{n=2}^{p-3} (n-1) \left( |F_{k+2}^{(n)_I}|^2 + \frac{p-6}{2} |F_{k+4}^{(n)_I}|^2 + \frac{p-7}{4} |F_5^{(n)_I}|^2 \right) \ , \nn
\eea
instead of \eqref{newR43}. There are new terms in $F_{k-2}$ and ${\cal P}_{IJ}$: they seem difficult to handle, even though they may simplify when setting to zero the BPS-like condition for $H$ and $F_{k-2}$. As discussed in Section \ref{sec:Minksolfinal}, this expression could be useful for Minkowski solutions with those fluxes turned-on.

\end{appendix}

\end{document}